\newcommand{\ONECOLUMN}{}

\ifdefined\ONECOLUMN
	\documentclass[review]{elsarticle}
\else
	\documentclass[10pt,twocolumn,twoside]{IEEEtran}
\fi

\usepackage{xcolor}
\usepackage{color}
\usepackage{amsmath}
\usepackage{amsthm}
\usepackage{amsfonts}
\usepackage{amssymb}
\usepackage{wasysym}
\usepackage{graphicx}
\usepackage{subcaption}

\usepackage{epsfig}

\usepackage{mathtools}
\DeclarePairedDelimiter{\floor}{\lfloor}{\rfloor}

\usepackage{tikz}
\usetikzlibrary{calc,positioning,intersections,decorations}
\usepackage{tkz-euclide}
\usetkzobj{all}

\newcommand{\dfn}{\triangleq}

\newcommand{\rw}{\rightarrow}

\newcommand{\mE}{{\mathcal E}}
\newcommand{\mF}{{\mathcal F}}

\newcommand{\mX}{{\mathcal X}}
\newcommand{\mP}{{\mathcal P}}
\newcommand{\mN}{{\mathcal N}}
\newcommand{\mR}{{\mathcal R}}
\newcommand{\mU}{{\mathcal U}}

\newcommand{\Real}{\mathbb{R}}
\newcommand{\modul}{~~{\sf mod}~~ }

\newtheorem{Teorema}{\bf Theorem}
\newtheorem{Lema}{\bf Lemma}

\newtheorem{Hipotesis}{\bf Assumption}
\newtheorem{Nota}{\bf Remark}
\newtheorem{Ejemplo}{\bf Example}

\newtheorem{Algoritmo}{\bf Algorithm}

\begin{document}

\begin{frontmatter}
	
\title{A proof of uniform convergence over time for a distributed particle filter}



%
%

\author{Joaqu\'in M\'iguez$^\dagger$}
\ead{j.miguez@qmul.ac.uk}

\author{Manuel A. V\'azquez$^\star$}
\ead{mvazquez@tsc.uc3m.es}

\address{$^\dagger$School of Mathematical Sciences, Queen Mary University of London.\\ Mile End Rd, E1 4NS London, UK.\\$^\star$Departamento de Teor\'{\i}a de la Se\~nal y Comunicaciones, Universidad Carlos III de Madrid, Avenida de la Universidad 30, 28911 Legan\'es, Madrid, Spain.}

\begin{abstract}

Distributed signal processing algorithms have become a hot topic during the past years. One class of algorithms that have received special attention are particles filters (PFs). However, most distributed PFs involve various heuristic or simplifying approximations and, as a consequence, classical convergence theorems for standard PFs do not hold for their distributed counterparts. In this paper, we analyze a distributed PF based on the non-proportional weight-allocation scheme of Bolic {\em et al} (2005) and prove rigorously that, under certain stability assumptions, its asymptotic convergence is guaranteed uniformly over time, in such a way that approximation errors can be kept bounded with a fixed computational budget. To illustrate the theoretical findings, we carry out computer simulations for a target tracking problem. The numerical results show that the distributed PF has a negligible performance loss (compared to a centralized filter) for this problem and enable us to empirically validate the key assumptions of the analysis.   
   
\end{abstract}

\end{frontmatter}

\section{Introduction}

Distributed signal processing algorithms have become a hot topic during the past years, propelled by fast technological developments in the fields of parallel computing, on one hand, and wireless sensor networks (WSNs), on the other. In parallel computing, algorithms are optimized to run fast on a set of concurrent processors (e.g., in a graphics processing unit (GPU) \cite{Suchard10}), while signal processing methods for WSNs are designed for their implementation over a collection of low-power nodes that communicate wirelessly and share the processing tasks \cite{Read14}.  Popular techniques in the WSN arena include consensus-based estimators \cite{Fagnani08,Kar09,Kar10}, diffusion-based adaptive algorithms \cite{Lopes08,Cattivelli10,Chen12} and distributed stochastic filters, including Kalman filters \cite{Schizas08a,Schizas08b} and particle filters (PFs) \cite{Hlinka12,Lee13,Dias13,Dias13b}. While consensus and diffusion algorithms require many iterations of message passing for convergence, PFs are {\em a priori} better suited for online estimation and prediction tasks. Unfortunately, most distributed PFs (DPFs) rely on simplifying approximations and their convergence cannot be guaranteed by the classical  theorems in \cite{Crisan01,DelMoral01c,Bain08}. One exception is the Markov chain distributed particle filter (MCDPF), for which analytical results exist \cite{Lee13}. However, the MCDPF converges asymptotically as sets of samples and weights are retransmitted repeatedly over the network according to a random scheme. From this point of view, it is as communication-intensive as consensus algorithms and, therefore, less appropriate for online processing compared to classical PFs.

The implementation of PFs on parallel computing systems has received considerable attention since these methods were originally proposed in \cite{Gordon93}. 
The efficient implementation of PFs on parallel devices such as GPUs and multi-core CPUs is not as straightforward as it seems {\em a priori} because these Monte Carlo algorithms involve a resampling step which is inherently hard to parallelize. This issue is directly addressed in \cite{Bolic05}, where two parallel implementations of the resampling step are proposed. While the approach of \cite{Bolic05} is sound, the authors focus on implementation issues and no proof of convergence of the resulting PFs is provided. Only very recently, a number of authors have proposed distributed particle filtering schemes with provable convergence \cite{Whiteley13,Verge13}. These methods have a fairly broad scope (the methodology in \cite{Whiteley13} can actually be seen as a generalization of the techniques in \cite{Bolic05}) yet they appear to be less suitable for practical implementations under communications or computing power constraints, as they involve considerable parallelization overhead \cite{Verge13} or depend on the centralized computation of certain statistics that involve the whole set of particles in the filter \cite{Whiteley13}. 

The goal of this paper is to provide a rigorous proof of convergence for a DPF that relies on the distributed resampling with non-proportional weight-allocation scheme of \cite{Bolic05} (later adapted for implementation over WSNs in \cite{Read14}). Under assumptions regarding the stability of  the state-space model underlying the PF, we prove that this algorithm converges asymptotically (as the number of particles generated by the filter increases) and uniformly over time. Time-uniform convergence implies that the estimation errors stay bounded without having to increase the computational effort of the filter over time. We provide explicit convergence rates for the DPF and discuss the implications of this result and the assumptions on which the analysis is based. The theoretical investigation is complemented by computer simulations of an indoor target tracking problem. For this specific system, we first show that the performance of the centralized and distributed PFs is very similar and then proceed to validate numerically a key assumption used in the analysis, related to the degree of cooperation among processing elements in the distributed computing system on which the algorithm is run. 


The rest of the paper is organized as follows. In Section \ref{sDPF} we describe the DPF of interest. In Section \ref{sAnalysis} we prove a uniform convergence result for this filter and discuss the implications of such result. Computer simulations are presented in Section \ref{sSimulations} and, finally, Section \ref{sConclusions} is devoted to the conclusions.

\section{A distributed particle filtering algorithm} \label{sDPF}

\subsection{State space systems and the standard particle filter} \label{ssStateSpaceModel}

The stochastic filtering problem consists in tracking the posterior probability distribution of the state variables of a random dynamic system. Often, the problem is restricted to the (broad) class of Markov state space systems with conditionally independent observations. Let $\{ X_n \}_{n \ge 0}$ denote the discrete-time random sequence of the system state variables, taking values on the $d_x$-dimensional set $\mX \subseteq \Real^{d_x}$, and let $\{ Y_n \}_{n \ge 1}$ denote the corresponding sequence of observations, taking values on $\Real^{d_y}$. The systems of interest are modeled by triplets of the form 
$
\left\{ \tau_0(dx), \quad \tau_n(dx|x_{n-1}), \quad g_n(y_n|x_n) \right\}_{n\ge 1}
$, where $\tau_0$ is the prior probability measure associated to the random variable (r.v.) $X_0$, $\tau_n(dx|x_{n-1})$ is a Markov kernel that determines the probability distribution of $X_n$ conditional on $X_{n-1}=x_{n-1}$, and $g_n(y_n|x_n)$ is the conditional probability density function (pdf) of the random observation $Y_n$, given the state $X_n=x_n$, with respect to (w.r.t.) the Lebesgue measure. The latter is most often used as the {\em likelihood} of $X_n=x_n$ given the observation $Y_n=y_n$. We write $g_n$ as a function of $x_n$ explicitly, namely $g_n^{y_n}(x_n) \dfn g_n(y_n|x_n)$, to emphasize this fact.

The goal in the stochastic filtering problem is to sequentially compute the posterior probability measures of $X_n$ given the observations $Y_{1:n}=y_{1:n}$, denoted $\pi_n(dx)$, for $n=0, 1, ...$ (note that $\pi_0=\tau_0$). Except for a few particular cases, e.g., the Kalman \cite{Kalman60,Anderson79} and Bene\v{s} \cite{Bain08} filters, $\pi_n$ cannot be computed exactly and numerical approximations are pursued instead. PFs are recursive Monte Carlo algorithms that generate random discrete approximations of the probability measures $\{ \pi_n; n \ge 1 \}$ \cite{Crisan01,DelMoral01c,Bain08}. At time $n$ a typical particle filtering algorithm produces a set of $N$ random samples (often termed {\em particles}) and associated importance weights, $\Omega_n = \{ x_n^{(i)}, w_n^{(i)*} \}_{i=1}^N$ with $W_n = \sum_{i=1}^N w_n^{(i)*}$, and approximate $\pi_n$ by way of the random probability measure $\pi_n^N=\frac{1}{W_n}\sum_{i=1}^N w_n^{(i)*} \delta_{x_n^{(i)}}$, where $\delta_x$ denotes the Dirac (unit) delta measure located at $x$.

It is common to analyze the convergence of PFs in terms of the approximation of integrals w.r.t. $\pi_n$ \cite{DelMoral00,Crisan01,Bain08,DelMoral01c,Miguez13b}. To be specific, let $f : \mX \rw \Real$ be a real function integrable w.r.t. $\pi_n$. Then we denote
$$
(f,\pi_n) \dfn \int f(x)\pi_n(dx)
$$
and approximate the latter integral (generally intractable) as
$$
(f,\pi_n) \approx (f,\pi_n^N) = \int f(x)\pi_n^N(dx) = \frac{1}{W_n}\sum_{i=1}^N w_n^{(i)*} f(x_n^{(i)}).
$$

\subsection{A distributed particle filter}

We describe a PF based on the distributed resampling with non-proportional allocation (DRNA) scheme of \cite[Section IV.A.3]{Bolic05} (see also \cite{Miguez07b,Balasingam11,Read14}). Assume that the set of weighted particles 
$
\Omega_n = \{ x_n^{(i)}, w_n^{(i)*} \}_{i=1}^N
$
can be split into $M$ disjoint sets, 
$$
\Omega_n^m = \{ x_n^{(m,k)}, w_{n}^{(m,k)*} \}_{k=1}^K, m=1, ..., M, \mbox{ such that } \Omega_n= \cup_{m=1}^M \Omega_n^m,
$$ 
each of them assigned to an independent processing element (PE). 
The total number of particles is $N=MK$, where $M$ is the number of PEs and $K$ is the number of particles per PE. At the $m$-th PE, $m=1, ..., M$, we additionally keep track of the aggregated weight 
$
W_n^{(m)*} = \sum_{k=1}^K w_n^{(m,k)*}
$ 
for that PE. 

Every $n_0$ time steps, the PEs exchange subsets of particles and weights by using some communication network \cite{Bolic05}. We formally represent this transfer of data among PEs by means of a deterministic one-to-one map 
$$\beta:\{ 1, ..., M \} \times \{ 1, ..., K \} \rw \{ 1, ..., M \} \times \{ 1, ..., K \}$$
that keeps the number of particles per PE, $K$, invariant. 
To be specific, $(u,v) = \beta(m,k)$ means that the $k$-th particle of the $m$-th PE is transmitted to the $u$-th PE, where it becomes particle number $v$. Typically, only subsets of particles are transmitted from one PE to a different one, hence $\beta(m,k)=(m,k)$ for many values of $k$ and $m$. 
The DPF of interest in this paper can be outlined as follows.

\begin{Algoritmo} \label{alDPF}
DPF based on the DRNA scheme of \cite[Section IV.A.3]{Bolic05}, with $M$ PEs, $K$ particles per PE and periodic particle exchanges every $n_0 $ time steps.
\begin{enumerate}
\item For $m=1, ..., M$ (concurrently) draw $x_0^{(m,k)} \sim \tau_0(dx)$, $k=1, ..., K$, and set $w_0^{(m,k)*} = \frac{1}{MK}$ and $W_0^{(m)*} = 1/M$.

\item Assume that $\{ x_{n-1}^{(m,k)}, w_{n-1}^{(m,k)*} \}_{k=1}^K$ and $W_{n-1}^{(m)*}$ are available for each $m=1, ..., M$. 
	\begin{enumerate}
	\item For $m=1, ..., M$ (concurrently) and $k=1, ..., K$,
	\begin{eqnarray}
	\mbox{draw } \bar x_n^{(m,k)} &\sim& \tau_n(dx|x_{n-1}^{(m,k)}), \nonumber \\
	\mbox{compute } \bar w_n^{(m,k)*} &=& w_{n-1}^{(m,k)*} g_n^{y_n}( \bar x_n^{(m,k)} ), \nonumber\\
	\mbox{and } \bar W_n^{(m)*} &=& \sum_{k=1}^{K} \bar w_n^{(m,k)*}. \nonumber
	\end{eqnarray}
	
	\item Local resampling: for $m=1, ..., M$ (concurrently) set $\tilde x_n^{(m,k)} = \bar x_n^{(m,j)}$ 
	with probability 
	$$\bar w_n^{(m,j)} = \bar w_n^{(m,j)*} / \bar W_n^{(m)*},
	$$ 
	for $k = 1, ..., K$ and $j \in \{ 1, ..., K \}$. 
	
	Set $\tilde w_n^{(m,k)*} = \bar W_n^{(m)*}/K$ for each $m$ and all $k$. 
	
	\item Particle exchange: If $n = rn_0$ for some $r \in \mathbb{N}$, then set 
	$$
	x_n^{\beta(m,k)} = \tilde x_n^{(m,k)}
	\quad \mbox{and} \quad
	w_n^{\beta(m,k)*} = \tilde w_n^{(m,k)*}
	$$ 
	for every $(m,k) \in \{ 1, ..., M \} \times \{ 1, ..., K \}$. Also set $W_n^{(m)*} = \sum_{k=1}^K w_n^{(m,k)*}$ for every $m=1, ..., M$.
	
	Otherwise, if $n \ne rn_0$, set $x_n^{(m,k)}=\tilde x_n^{(m,k)}$, $w_n^{(m,k)*} = \tilde w_n^{(m,k)*}$, $W_n^{(m)*} = \bar W_n^{(m)*}$.
	\end{enumerate}
\end{enumerate}
\end{Algoritmo}

Every PE operates independently of all others except for the particle exchange, step 2.c), which is performed every $n_0$ time steps. The degree of interaction can be controlled by designing the map $\beta(m,k)$ in a proper way. Typically, exchanging a subset of particles with ``neighbor'' PEs is sufficient, as illustrated by the following example.

\begin{Ejemplo}
Consider a circular arrangement in which the $m$-th PE exchanges particles with PE $(m-1) \modul M$ and $(m+1) \modul M$, where $\modul$ indicates the modulus operation ($a \modul b$ is the integer remainder of the division $a/b$). To be explicit,  
\begin{itemize}
\item for each $m = 2, ..., M-1$, the $m$-th PE exchanges particles with two neighbours, namely PE number $(m-1)$ and PE number $(m+1)$,
\item PE number 1 exchanges particles with PE number $M$ and PE number 2, and
\item PE number $M$ exchanges particles with PE number $M-1$ and PE number 1.
\end{itemize}

Next, assume for simplicity that each PE sends one particle to each one of its neighbors (i.e., it sends out {\em two} particles) and receives one particle from each one of its neighbors as well (i.e., it gets {\em two} new particles) so that the number of particles per PE, $K$ remains constant. One choice of map $\beta$ that implements such an exchange is the following
$$
\beta(m,k) := \left\{
	\begin{array}{ll}
	(m,k) &\mbox{if $2 \le k \le K-1$,}\\
	(m-1 \modul M, K) &\mbox{if $k=1$,}\\
	(m+1 \modul M,1) &\mbox{if $k=K$.}\\
	\end{array}
\right.
$$
In other words, for each PE $m =1, ..., M$,
\begin{itemize}
\item all particles with index $k=2, ..., K-1$ remain the same,
\item the first particle ($k=1$) is sent to PE $m-1 \modul M$ and, in exchange, the $K$-th particle from that neighbor is received, i.e., $x_n^{(m,1)} = \tilde x_n^{(m-1 \modul M,K)}$ and $w_n^{(m,1)*} = \tilde w_n^{(m-1 \modul M,K)*}$,
\item  the last particle ($k=K$) is sent to PE $m+1 \modul M$ and, in exchange, the $1$-st particle from that neighbor is received, i.e., $x_n^{(m,K)} = \tilde x_n^{(m+1 \modul M,1)}$ and $w_n^{(m,K)*} = \tilde w_n^{(m+1 \modul M,1)*}$.
\end{itemize}

It is apparent that this instance of $\beta$ preserves the number of particles per PE $K$ constant. More elaborate schemes can be designed and in general this is related to the structure of the communication network that interconnects the PEs. As long as it is guaranteed that the number of particles that a PE gives to its neighbors is the same as the number of particles that it receives from these neighbors, the number $K$ of particles per PE remains invariant.
\end{Ejemplo}

\begin{Nota} \label{rmWresampling}
The local resampling step 2.b) is carried out independently, and concurrently, at each PE and it does not change the aggregate weights, i.e., 
$\bar W_n^{(m)*} = \sum_{k=1}^K \bar w_n^{(m,k)*} = \sum_{k=1}^K \tilde w^{(m,k)*}$. We assume a multinomial resampling procedure, but other schemes (see, e.g., \cite{Bain08}) can be easily incorporated.
\end{Nota}

\subsection{Measure and integral approximations}

Let 
\ifdefined\ONECOLUMN
	$$
	\bar w_n^{(m,k)} = \frac{
		\bar w_n^{(m,k)*} 
	}{
		\sum_{i=1}^K \bar w_n^{(m,i)*}
	}, \quad \tilde w_n^{(m,k)} = \frac{
		\tilde w_n^{(m,k)*} 
	}{
		\sum_{i=1}^K \tilde w_n^{(m,i)*}
	} \quad \mbox{and} \quad
	w_n^{(m,k)} = \frac{
		w_n^{(m,k)*} 
	}{
		\sum_{i=1}^K w_n^{(m,i)*}
	}
	$$
\else
$$
	\begin{array}{cc}
		\bar w_n^{(m,k)} = \frac{
			\bar w_n^{(m,k)*} 
		}{
			\sum_{i=1}^K \bar w_n^{(m,i)*}
		}
		, &
		\tilde w_n^{(m,k)} = \frac{
			\tilde w_n^{(m,k)*} 
		}{
			\sum_{i=1}^K \tilde w_n^{(m,i)*}
		}
		,\\{}\\
		\multicolumn{2}{c}
		{
			\text{and }
			w_n^{(m,k)} = \frac{
				w_n^{(m,k)*} 
			}{
				\sum_{i=1}^K w_n^{(m,i)*}
			}
		}
	\end{array}
$$
\fi
be the locally normalized versions of the importance weights and let 
\ifdefined\ONECOLUMN
	$$
	\bar W_n^{(m)} = \frac{
		\bar W_n^{(m)*}
	}{
		\sum_{i=1}^M \bar W_n^{(i)*}
	} \quad \mbox{and} \quad W_n^{(m)} = \frac{
		W_n^{(m)*}
	}{
		\sum_{i=1}^M W_n^{(i)*}
	}
	$$
\else
	$$
		\begin{array}{ccc}
			\bar W_n^{(m)} = \frac{
				\bar W_n^{(m)*}
			}{
				\sum_{i=1}^M \bar W_n^{(i)*}
			}
			& \text{and} &
			W_n^{(m)} = \frac{
				W_n^{(m)*}
			}{
				\sum_{i=1}^M W_n^{(i)*}
			}
		\end{array}	
	$$
\fi
be the globally normalized aggregated weights of the $m$-th PE before and after the particle exchange step, respectively. The DPF produces three different local approximations of the posterior measure $\pi_n$ at each PE, namely
\ifdefined\ONECOLUMN
	$$
	\bar \pi_n^{m,K} = \sum_{k=1}^K \bar w_n^{(m,k)} \delta_{\bar x_n^{(m,k)}}, \quad
	\tilde \pi_n^{m,K} =  \frac{1}{K} \sum_{k=1}^K \delta_{\tilde x_n^{(m,k)}} \quad 
	\mbox{and} \quad
	\pi_n^{m,K} = \sum_{k=1}^K w_n^{(m,k)} \delta_{x_n^{(m,k)}},
	$$
\else
	$$
	\begin{array}{cc}
		\bar \pi_n^{m,K} = \sum_{k=1}^K \bar w_n^{(m,k)} \delta_{\bar x_n^{(m,k)}}
		,&
		\tilde \pi_n^{m,K} =  \frac{1}{K} \sum_{k=1}^K \delta_{\tilde x_n^{(m,k)}}
		,\\{}\\
		\multicolumn{2}{c}
		{
			\text{and }
			\pi_n^{m,K} = \sum_{k=1}^K w_n^{(m,k)} \delta_{x_n^{(m,k)}}
		}
	\end{array}
	$$
\fi
corresponding to steps 2.a), 2.b) and 2.c) of Algorithm \ref{alDPF}. The normalized aggregate weights can be used to combine the local approximations, which readily yields global approximations of the posterior measure, i.e., 
\ifdefined\ONECOLUMN
	\begin{equation}
	\bar \pi_n^{MK} = \sum_{m=1}^M \bar W_n^{(m)} \bar \pi_n^{m,K}, \quad
	\tilde \pi_n^{MK} = \sum_{m=1}^M \bar W_n^{(m)} \tilde \pi_n^{m,K} \quad \mbox{and} \quad
	\pi_n^{MK} = \sum_{m=1}^M W_n^{(m)} \pi_n^{m,K}.
	\label{eqPiN}
	\end{equation}
\else
	\begin{equation}
		\begin{array}{cc}
			\bar \pi_n^{MK} = \sum_{m=1}^M \bar W_n^{(m)} \bar \pi_n^{m,K}
			,&
			\tilde \pi_n^{MK} = \sum_{m=1}^M \bar W_n^{(m)} \tilde \pi_n^{m,K}
			,\\{}\\
			\multicolumn{2}{c}
			{
				\text{and }
				\pi_n^{MK} = \sum_{m=1}^M W_n^{(m)} \pi_n^{m,K}
			}
		\end{array}
		.
		\label{eqPiN}
	\end{equation}
\fi
Note that only the local normalization of the weights $\bar w_n^{(m,k)}$, $k=1, ..., K$, is necessary for Algorithm \ref{alDPF} to run (as they are needed in the local resampling step 2.b). The computation of the $w_n^{(m,k)}$'s, the $W_n^{(m)}$'s or $\pi_n^{MK}$ are only necessary when local or global estimates of $\pi_n$ are needed. However, the computation of these estimates can be carried out concurrently with Algorithm \ref{alDPF}, i.e., the DPF can keep running in parallel with the computation of any estimates.  

\begin{Nota} \label{rmMiscelanea}
Algorithm \ref{alDPF} enjoys some relatively straightforward properties that should be highlighted, as they are relevant for the analysis in Section \ref{sAnalysis}.
\begin{enumerate}
\item The particle exchange step does not change the particles or their weights. It only ``shuffles'' the particles among the PEs and updates the aggregate weights accordingly. As a result, 
$\pi_n^{MK} = \tilde \pi_n^{MK}$, since the individual particles and weights are not changed.

\item A random exchange (i.e., a random $\beta$) is also possible, although it makes certain practical implementations harder \cite{Bolic05}. We abide by a deterministic map $\beta$ for the sake of conciseness, although the analysis can be extended to account for random schemes in a relatively straightforward manner. 

\item The ensemble of local resampling steps keeps the local and aggregate importance weights proper and is globally unbiased \cite{Read14}.

\end{enumerate}
\end{Nota}

The goal of this paper is to analyze the approximation of integrals using the random measure $\pi_n^{MK}$ in \eqref{eqPiN}. 
We look into the $L_p$ norm of the approximation errors, namely
$
\| (f,\pi_n^{MK}) - (f,\pi_n) \|_p
$ 
for $p \ge 1$, where $f:\mX \rw \Real$ is an integrable real function on $\mX$ and $\| \cdot \|_p = E[|\cdot|^p]^\frac{1}{p}$. The expectation is taken w.r.t. the distribution of the r.v. $\pi_n^{MK}$.

\section{Analysis} \label{sAnalysis}

\subsection{Assumptions, preliminary results and notations}

Let $\mP(\mX)$ be the set of probability measures on $\left(\mathcal{B}(\mX),\mX\right)$, where $\mathcal{B}(\mX)$ is the Borel $\sigma$-algebra of open subsets of the state space $\mX$. Choose a measure $\alpha \in \mP(\mX)$ and let $h : \mX \rw \Real$ be a real function integrable w.r.t. $\alpha$. We define the measure-valued map $\Psi_n : \mP(\mX) \rw \mP(\mX)$ as
$$
(h,\Psi_n(\alpha)) \dfn \frac{
	\left(
		(hg_n^{y_n},\tau_n), \alpha
	\right)
}{
	\left(
		(g_n^{y_n},\tau_n), \alpha
	\right)
}
$$
and it is not difficult to show that $\Psi_n$ is the transformation that converts the filter measure at time $n-1$ into the filter at time $n$, i.e., $\pi_n = \Psi_n(\pi_{n-1})$ \cite{DelMoral01c,Bain08}. Composition of maps is denoted 
$
\Psi_{n|s} \dfn \Psi_n \circ \Psi_{n-1} \circ \cdots \circ \Psi_{s+1}
$, for $s < n$, hence $\Psi_{n|n-1}(\alpha)=\Psi_n(\alpha)$, and we adopt the convention $\Psi_{n|n}(\alpha)=\alpha$. 
We also define the functional $\Gamma_{n|s} : \left( 
	\mX \rw \Real
\right) \rw \left(
	\mX \rw \Real
\right)$, recursively, as
\begin{eqnarray}
\Gamma_{n|n}(h) &\dfn& h, \nonumber\\
\Gamma_{n|n-r}(h) &\dfn& \left(
	g_{n-r+1}^{y_{n-r+1}} \Gamma_{n|n-r+1}(h), \tau_{n-r+1}
\right), \quad r \ge 1,\nonumber
\end{eqnarray}
for $s \le n$, and it is not difficult to show that \cite{DelMoral01c}
\begin{equation}
\left( h, \Psi_{n|n-r}(\alpha) \right) = \frac{
	\left( \Gamma_{n|n-r}(h), \alpha \right)
}{
	\left( \Gamma_{n|n-r}(1), \alpha \right)
},
\label{eqPsiGamma}
\end{equation}
where $1(x)=1$ is the constant unit function.

For conciseness, we denote the set of bounded real functions over $\mX$ as $B(\mX)$, i.e., $h \in B(\mX)$ if, and only if, $h$ is a map of the form $\mX \rw \Real$ and $\| h \|_\infty = \sup_{x \in \mX} | h(x) | < \infty$. In the sequel, we analyze the asymptotics of the approximation errors $|(h,\pi_n^N)-(h,\pi_n)|$, where $h \in B(\mX)$, subject to the following assumptions.

\begin{Hipotesis} \label{asBounded}
There exists a bounded sequence of positive real numbers $\{ a_n \}_{n \ge 0}$ such that $\frac{1}{a_n} \le g_n^{y_n}(x) \le a_n$, for every $x \in \mX$, and $a = \sup_{n \ge 0} a_n < \infty$.
\end{Hipotesis}
\begin{Hipotesis} \label{asStable}
For any probability measures $\alpha, \eta \in \mP(\mX)$ and every $h \in B(\mX)$,
$$
\mE(h,T) = \sup_{n \ge 0} \left|
	(h,\Psi_{n+T|n}(\alpha)) - (h,\Psi_{n+T|n}(\eta))
\right| < \infty,
 \quad \mbox{and} \quad
\lim_{T\rw\infty} \mE(h,T) = 0.
$$
\end{Hipotesis}

Assumption \ref{asBounded} states that the likelihood functions are upper-bounded as well as bounded away from 0. Assumption \ref{asStable} states that the optimal filter $\pi_n$ for the given state-space system is stable. 
A detailed study of the stability properties of optimal filters for the class of state space models of interest here can be found in \cite{DelMoral01c} (see also \cite{VanHandel09,VanHandel09b} for recent developments), including conditions on the kernels $\tau_n$ and the likelihoods $g_n^{y_n}$ which are sufficient to ensure stability.

\begin{Hipotesis} \label{asExchange}
The particle exchange step, with period $n_0$, guarantees that 
\begin{equation}
E\left[
	\left(
		\sup_{1 \le m \le M} W_{rn_0}^{(m)} 
	\right)^q
\right] \le \frac{c^q}{M^{q-\epsilon}}, \quad \mbox{for every $r \in \mathbb{N}$}
\label{eqIneq-ass3}
\end{equation}
and some constants $c < \infty$, $0 \le \epsilon < 1$ and $q \ge 4$ independent of $M$.
\end{Hipotesis}




Intuitively, Assumption \ref{asExchange} says that the aggregate weights remain ``sufficiently balanced'' (i.e., no PE takes too much weight compared to others). We also introduce the lemma below which, combined with Assumption \ref{asExchange}, is key to the analysis of the approximation errors, as it enables us to obtain tractable bounds for the aggregate weights. 

\begin{Lema} \label{lmW}
Assume that
\begin{equation}
E\left[
        \left( 
        		\sup_{1 \le m \le M} W_n^{(m)} 
	\right)^q
\right] \le \frac{
        c^q
}{
        M^{q-\epsilon}
}, \label{eqAss_lema_W}
\end{equation}
for some $q \ge 2$, $c>0$ and $0 \le \epsilon < 1$ constant w.r.t. $M$. Then, there exists a non-negative and a.s. finite random variable $U_n^\varepsilon$, independent of $M$, such that
\begin{equation}
\sup_{1 \le m \le M} W_n^{(m)} \le \frac{
        U_n^\varepsilon
}{
        M^{1-\varepsilon}
}, 
\end{equation}
where $\frac{1+\epsilon}{q}<\varepsilon<1$ is also a constant w.r.t. $M$. Moreover, there is another constant $u^{\varepsilon,q} < \infty$ independent of $n$ and $M$ such that
$
\sup_{n\ge 0} E\left[ \left( U_n^\varepsilon \right)^q \right] < u^{\varepsilon,q}.
$
\end{Lema}

\noindent {\bf Proof.} See \ref{apLemmaW}. $\qed$

\begin{Nota} \label{rmMeasurable}
The r.v. $U_n^\varepsilon$ can be written as
$$
U_n^\varepsilon = \sum_{M=1}^\infty M^{q-1-\gamma} \left( \sup_{1 \le m \le M} W_n^{(m)} \right)^q
$$
(see Eq. \eqref{eqShow_it_is_measurable} in \ref{apLemmaW}), where $\epsilon < \gamma < q - 1$ is constant w.r.t. $M$. If we also note that the aggregate weights after the particle exchange step, $\{ W_n^{(m)*} \}_{m=1}^M$, can be computed deterministically\footnote{Because the map $\beta(m,k)$ used for the particle exchange is deterministic.} 
given the aggregates before the exchange, $\{ \bar W_n^{(m)*} \}_{m=1}^M$, then it follows that $U_n^\varepsilon$ is measurable w.r.t. the $\sigma$-algebra $\bar \mF_n^\infty = \bigcup_{M\ge 1} \bar \mF_n^M$, where each term in the countable union is a generated $\sigma$-algebra, namely
$
\bar \mF_n^M = \sigma\left( 
	x_{0:n-1}^{(m,j)}, \bar x_{0:n}^{(m,j)}; \quad 1 \le m \le M, \quad 1 \le j \le K 
\right).
$
\end{Nota}

Finally, we introduce a simple inequality that will be repeatedly used through the analysis of Algorithm \ref{alDPF}. Let $\alpha, \beta, \bar \alpha, \bar \beta \in \mP(\mX)$ be probability measures and let $f,h \in B(\mX)$ be two real bounded functions on $\mX$ such that $(h,\bar \alpha)>0$ and $(h,\bar \beta)>0$. If the identities 
$
(f,\alpha) = \frac{
	(fh,\bar \alpha)
}{
	(h,\bar \alpha)
} \quad \mbox{and} \quad (f,\beta) = \frac{
	(fh,\bar \beta)
}{
	(h,\bar \beta)
}
$
hold, then it is straightforward to show (see, e.g., \cite{Crisan01}) that
\ifdefined\ONECOLUMN
	\begin{equation}
	| (f,\alpha)-(f,\beta) | \le \frac{
		1
	}{
		(h,\bar \alpha)
	} \left|
		(fh,\bar \alpha) - (fh,\bar \beta)
	\right| + \frac{
		\| f \|_\infty
	}{
		(h,\bar \alpha)
	} \left|
		(h,\bar \alpha) - (h,\bar \beta)
	\right|. 
	\label{eqPreliminaries}
	\end{equation}
\else
	\begin{align}
		| (f,\alpha)-(f,\beta) | 
		\le {} &
		\frac{
			1
		}{
			(h,\bar \alpha)
		} \left|
			(fh,\bar \alpha) - (fh,\bar \beta)
		\right| -
		\nonumber\\ & -
		\frac{
			\| f \|_\infty
		}{
			(h,\bar \alpha)
		} \left|
			(h,\bar \alpha) - (h,\bar \beta)
		\right|. 
		\label{eqPreliminaries}
	\end{align}
\fi


\subsection{Uniform convergence over time}

In this section we rigorously prove that $\| (h,\pi_n^{MK}) - (h,\pi_n) \|_p \rw 0$, as $M \rw \infty$ and $K$ remains fixed, uniformly over time. The key result is Lemma \ref{lmPsi} below, on the propagation of errors across the map $\Psi_n$. From this result, we then obtain the main theorem on the convergence of Algorithm \ref{alDPF}.

\begin{Lema} \label{lmPsi}
Let $K < \infty$ be fixed. If Assumption \ref{asBounded} holds, with $a < \infty$, and Assumption \ref{asExchange} holds, with $0 < \epsilon < 1$ and $q \ge 4$, then there exist constants $u^\varepsilon < \infty$ and $\tilde c < \infty$, independent of $M$, such that
\ifdefined\ONECOLUMN
	\begin{equation}
	\| (h,\Psi_{n|n-k}(\pi_{n-k}^{MK})) - (h,\Psi_{n|n-k-1}(\pi_{n-k-1}^{MK})) \|_p \le \frac{
		8a^{k+2} u^\varepsilon \tilde c \| h \|_\infty
	}{
		M^{\frac{1}{2}-\varepsilon} K^\frac{1}{2}
	}, \label{eqX10}
	\end{equation}
\else
	\begin{align}
		& \| (h,\Psi_{n|n-k}(\pi_{n-k}^{MK})) - (h,\Psi_{n|n-k-1}(\pi_{n-k-1}^{MK})) \|_p
		\le \nonumber\\ & \le
		\frac{
			8a^{k+2} u^\varepsilon \tilde c \| h \|_\infty
		}{
			M^{\frac{1}{2}-\varepsilon} K^\frac{1}{2}
		}, \label{eqX10}
	\end{align}
\fi
for every $h \in B(\mX)$, $k \le n$, $p \le q$, and $\varepsilon \in \left( \frac{1+\epsilon}{q}, \frac{1}{2} \right)$.
\end{Lema}

\noindent {\bf Proof.} Let us write $\pi_n^N = \pi_n^{MK}$ in the remaining of the proof for conciseness. We can use the relationship \eqref{eqPsiGamma} to rewrite the $L_p$ norm of the approximation error $(h,\Psi_{n|n-k}(\pi_{n-k}^N)) - (h,\Psi_{n|n-k-1}(\pi_{n-k-1}^N))$ as 
\ifdefined\ONECOLUMN
	\begin{align*}
		& \| (h,\Psi_{n|n-k}(\pi_{n-k}^N)) - (h,\Psi_{n|n-k-1}(\pi_{n-k-1}^N)) \|_p
		\le \nonumber\\ & \le
		\left\|
				\frac{
					(\Gamma_{n|n-k}(h), \pi_{n-k}^N)
				}{
					(\Gamma_{n|n-k}(1), \pi_{n-k}^N)
				} - \frac{
					(\Gamma_{n|n-k}(h), \Psi_{n-k}(\pi_{n-k-1}^N))
				}{
					(\Gamma_{n|n-k}(1), \Psi_{n-k}(\pi_{n-k-1}^N))
				} 
		\right\|_p
	\end{align*}
\else
	\begin{align*}
		& \| (h,\Psi_{n|n-k}(\pi_{n-k}^N)) - (h,\Psi_{n|n-k-1}(\pi_{n-k-1}^N)) \|_p
		\le \nonumber\\ & \le
		\left\|
				\frac{
					(\Gamma_{n|n-k}(h), \pi_{n-k}^N)
				}{
					(\Gamma_{n|n-k}(1), \pi_{n-k}^N)
				} - \frac{
					(\Gamma_{n|n-k}(h), \Psi_{n-k}(\pi_{n-k-1}^N))
				}{
					(\Gamma_{n|n-k}(1), \Psi_{n-k}(\pi_{n-k-1}^N))
				} 
		\right\|_p
	\end{align*}
\fi
and applying \eqref{eqPreliminaries} together with Assumption \ref{asBounded} in the equation above, we readily find an upper bound of the form
\ifdefined\ONECOLUMN
	\begin{equation}
	\| (h,\Psi_{n|n-k}(\pi_{n-k}^N)) - (h,\Psi_{n|n-k-1}(\pi_{n-k-1}^N)) \|_p \le a^k \left[ 
		\| e_{n,k}^N(h) \|_p + \| h \|_\infty \| e_{n,k}^N(1) \|_p
	\right], \label{eqIneq2}
	\end{equation}
\else
	\begin{align}
		& \| (h,\Psi_{n|n-k}(\pi_{n-k}^N)) - (h,\Psi_{n|n-k-1}(\pi_{n-k-1}^N)) \|_p
		\le \nonumber\\ & \le
		a^k \left[ 
			\| e_{n,k}^N(h) \|_p + \| h \|_\infty \| e_{n,k}^N(1) \|_p
		\right]
		\label{eqIneq2}
		,
	\end{align}
\fi
where 
\begin{equation}
e_{n,k}^N(h) = (\Gamma_{n|n-k}(h), \pi_{n-k}^N) - (\Gamma_{n|n-k}(h), \Psi_{n-k}(\pi_{n-k-1}^N))
\label{eqBoudForE}
\end{equation}
(note that $\Gamma_{n|n-k}(1) \ge a^{-k}$). 

The two terms between square brackets on the right hand side (rhs) of \eqref{eqIneq2} have the same form. To upper-bound them, we need to find bounds for errors of the form 
$
\| (v, \pi_{n-k}^N) - (v, \Psi_{n-k}(\pi_{n-k-1}^N)) \|_p
$, where $v \in B(\mX)$. To do this, we first split the $L_p$ norm of the error using a triangle inequality,
\ifdefined\ONECOLUMN
	\begin{eqnarray}
		\| (v, \pi_{n-k}^N) - (v, \Psi_{n-k}(\pi_{n-k-1}^N)) \|_p &\le& \| (v, \pi_{n-k}^N) - (v,\bar \pi_{n-k}^N) \|_p \label{eqIneq2.25} \\
		&& + \| (v, \bar \pi_{n-k}^N) - (v,\Psi_{n-k}(\pi_{n-k-1}^N)) \|_p.\nonumber
	\end{eqnarray}
\else
	\begin{align}
		& \| (v, \pi_{n-k}^N) - (v, \Psi_{n-k}(\pi_{n-k-1}^N)) \|_p
		\le \| (v, \pi_{n-k}^N)
		- \nonumber\\ & 
		- (v,\bar \pi_{n-k}^N) \|_p +
		\| (v, \bar \pi_{n-k}^N) - (v,\Psi_{n-k}(\pi_{n-k-1}^N)) \|_p
		.
		\label{eqIneq2.25} 
	\end{align}
\fi
To deduce a bound for the first term on the rhs of \eqref{eqIneq2.25}, let us 
recall that the particle exchange step does not modify the individual particle weights, only the aggregates, hence $\pi_n^N = \tilde \pi_n^N$. Then, we can readily write the conditional expectation of $|(v, \pi_{n-k}^N) - (v,\bar \pi_{n-k}^N)|^p$  (given $\bar \mF_{n-k}^\infty$, see Remark \ref{rmMeasurable}) as
\ifdefined\ONECOLUMN
	\begin{eqnarray}
	E\left[
		\left|
			(v, \pi_{n-k}^N) - (v,\bar \pi_{n-k}^N) 
		\right|^p | \bar \mF_{n-k}^\infty
	\right] &=& E\left[
		\left|
			(v, \tilde \pi_{n-k}^N) - (v,\bar \pi_{n-k}^N) 
		\right|^p | \bar \mF_{n-k}^\infty
	\right] \nonumber\\
	&=& E\left[
		\left|  
			\sum_{m=1}^M \frac{\bar W_{n-k}^{(m)}}{K} \sum_{j=1}^K \bar Z_{n-k}^{(m,j)}
		\right|^p | \bar \mF_{n-k}^\infty
	\right], \quad  \label{eqEq2.5}
	\end{eqnarray}
\else
	\begin{align}
		&
		E\left[
			\left|
				(v, \pi_{n-k}^N) - (v,\bar \pi_{n-k}^N) 
			\right|^p | \bar \mF_{n-k}
		\right]
		= \nonumber\\ & =
		E\left[
			\left|
				(v, \tilde \pi_{n-k}^N) - (v,\bar \pi_{n-k}^N) 
			\right|^p | \bar \mF_{n-k}
		\right]
		\nonumber\\ & =
		E\left[
			\left|  
				\sum_{m=1}^M \frac{\bar W_{n-k}^{(m)}}{K} \sum_{j=1}^K \bar Z_{n-k}^{(m,j)}
			\right|^p | \bar \mF_{n-k}
		\right]
		,
		\label{eqEq2.5}
	\end{align}
\fi
where the r.v.'s $\bar Z_{n-k}^{(m,j)} = v(\tilde x_{n-k}^{(m,j)}) - (v, \bar \pi_{n-k}^N)$ are conditionally independent (given $\bar \mF_{n-k}^\infty$),
zero mean (since $E[ v(\tilde x_{n-k}^{(m,j)}) | \bar \mF_{n-k}^\infty ] = (v,\bar \pi_{n-k}^N)$) and bounded (namely, $\| \bar Z_{n-k}^{(m,j)} \|_\infty \le 2\| v \|_\infty$ for all $n$, $k$, $m$ and $j$). Additionally, from step 2.a) of Algorithm \ref{alDPF} it follows that the normalized aggregate weights $\bar W_{n-k}^{(m)}$, $m=1, ..., M$, have the form 
\begin{eqnarray}
\bar W_{n-k}^{(m)} &=& \frac{
	\sum_{j=1}^K g_{n-k}^{y_{n-k}}(\bar x_{n-k}^{(m,j)}) w_{n-k-1}^{(m,j)*}
}{
	\sum_{l=1}^M  \sum_{i=1}^K g_{n-k}^{y_{n-k}}(\bar x_{n-k}^{(l,i)}) w_{n-k-1}^{(l,i)*}
}  \le  \frac{
	a^2 \sum_{j=1}^K w_{n-k-1}^{(m,j)*}
}{
	\sum_{l=1}^M  \sum_{i=1}^K w_{n-k-1}^{(l,i)*}
} \label{eqIneqW} \\
&=&  a^2 W_{n-k-1}^{(m)} 
\label{eqIneq2.9}
\end{eqnarray} 
where the inequality \eqref{eqIneqW} is a consequence of Assumption \ref{asBounded}, while \eqref{eqIneq2.9} follows immediately from the definition of the weights in Algorithm \ref{alDPF}. However, given \eqref{eqIneq2.9} and provided that there is no particle exchange at times $n-k-1, n-k-2, ...$ (exchanges occur periodically with period $n_0$) we readily obtain a straightforward relationship in the sequence of aggregate weights, namely
\begin{equation}
\bar W_{n-k}^{(m)} \le 
a^2 W_{n-k-1}^{(m)}  = 
a^2 \bar W_{n-k-1}^{(m)} \le 
a^4 W_{n-k-2}^{(m)} =
a^4 \bar W_{n-k-2}^{(m)} \le \cdots
\label{eqIneq3}
\end{equation}
Since the most recent particle exchange was carried out at most $n_0$ time steps earlier, we can readily iterate \eqref{eqIneq3} to obtain
\ifdefined\ONECOLUMN
	\begin{eqnarray}
	\bar W_{n-k}^{(m)} &\le& a^2 \bar W_{n-k-1}^{(m)} 
	\le \cdots \le a^{2\left( n - k - n_0\lfloor (n-k)/n_0 \rfloor \right)} W_{n_0 \lfloor (n-k)/n_0 \rfloor}^{(m)} \nonumber \\
	&\le& a^{2n_0} W_{n_0\lfloor (n-k)/n_0 \rfloor}^{(m)}.
	\label{eqIneq4}
	\end{eqnarray}
\else
	\begin{align}
		\bar W_{n-k}^{(m)}
		& \le
		a^2 \bar W_{n-k-1}^{(m)}
		\le \cdots \le
		a^{2\left( n - k - n_0\lfloor \frac{n-k}{n_0} \rfloor \right)} W_{n_0 \lfloor \frac{n-k}{n_0} \rfloor}^{(m)}
		\nonumber\\ & \le
		a^{2n_0} W_{n_0\lfloor \frac{n-k}{n_0} \rfloor}^{(m)}
		.
		\label{eqIneq4}
	\end{align}
\fi
However, the inequality \eqref{eqIneq4} combined with Assumption \ref{asExchange} yields
\ifdefined\ONECOLUMN
	\begin{equation}
	E\left[
		\left( \sup_{1 \le m \le M} \bar W_{n-k}^{(m)} \right)^q
	\right] \le a^{2n_0q} E\left[
		\left( \sup_{1 \le m \le M} W_{n_0\lfloor (n-k)/n_0 \rfloor}^{(m)} \right)^q
	\right] \le \frac{
		( a^{2n_0} c )^q
	}{
		M^{q-\epsilon}
	}
	\label{eqX1}
	\end{equation}
\else
	\begin{align}
		E\left[
			\left( \sup_{1 \le m \le M} \bar W_{n-k}^{(m)} \right)^q
		\right]
		& \le
		a^{2n_0q} E\left[
			\left( \sup_{1 \le m \le M} W_{n_0\lfloor \frac{n-k}{n_0} \rfloor}^{(m)} \right)^q
		\right]
		\nonumber\\ & \le
		\frac{
			( a^{2n_0} c )^q
		}{
			M^{q-\epsilon}
		}
		\label{eqX1}
	\end{align}
\fi
for some $q \ge 4$, where $c>0$, $n_0 \ge 1$ and $0 \le \epsilon < 1$ are constants independent of $M$, $K$ and $n$. In turn, the inequality \eqref{eqX1} enables the application of Lemma \ref{lmW}, which states that there exists an a.s. finite r.v. $U_{n-k}^\varepsilon$, independent of $M$, such that
\begin{equation}
\sup_{1 \le m \le M} \bar W_{n-k}^{(m)} \le \frac{
	U_{n-k}^\varepsilon
}{
	M^{1-\varepsilon}
},
\label{eqX2}
\end{equation}
where $\frac{1+\epsilon}{q} < \varepsilon < \frac{1}{2}$ is also constant w.r.t. $M$. Substituting \eqref{eqX2} back into Eq. \eqref{eqEq2.5} we arrive at
\ifdefined\ONECOLUMN
	\begin{equation}
	E\left[
		\left|
			(v, \pi_{n-k}^N) - (v,\bar \pi_{n-k}^N) 
		\right|^p | \bar \mF_{n-k}^\infty
	\right] \le E\left[
		\left|
			\frac{
				U_{n-k}^\varepsilon M^\varepsilon
			}{
				MK
			} \sum_{m=1}^M \sum_{j=1}^K \bar Z_{n-k}^{(m,j)}
		\right|^p | \bar \mF_{n-k}^\infty
	\right].
	\label{eqX3}
	\end{equation}
\else
	\begin{align}
		&
		E\left[
			\left|
				(v, \pi_{n-k}^N) - (v,\bar \pi_{n-k}^N) 
			\right|^p | \bar \mF_{n-k}^\infty
		\right]
		\le \nonumber\\& \le
		E\left[
			\left|
				\frac{
					U_{n-k}^\varepsilon M^\varepsilon
				}{
					MK
				} \sum_{m=1}^M \sum_{j=1}^K \bar Z_{n-k}^{(m,j)}
			\right|^p | \bar \mF_{n-k}^\infty
		\right]
		.
		\label{eqX3}
	\end{align}
\fi
Since $U_{n-k}^\varepsilon$ is measurable w.r.t. $\bar \mF_{n-k}^\infty$ (see Remark \ref{rmMeasurable}) and the r.v.'s $\bar Z_{n-k}^{(m,j)}$ are conditionally independent, have zero mean and upper bound $2\| v \|_\infty$, it is an exercise in combinatorics to show that
\begin{equation}
E\left[
	\left|
		(v, \pi_{n-k}^N) - (v,\bar \pi_{n-k}^N) 
	\right|^p | \bar \mF_{n-k}^\infty
\right] \le 
\frac{
	(2 U_{n-k}^\varepsilon M^\varepsilon \tilde c \| v \|_\infty)^p
}{
	(MK)^\frac{p}{2}
}
\label{eqX4}
\end{equation}
for some constant $\tilde c < \infty$ independent of $n$, $M$ and $K$ (actually, independent of the distribution of the $\bar Z_{n-k}^{(m,j)}$'s). Taking unconditional expectations on both sides of the inequality in \eqref{eqX4} yields 
\begin{equation}
E\left[
	\left|
		(v, \pi_{n-k}^N) - (v,\bar \pi_{n-k}^N) 
	\right|^p
\right] \le E\left[ 
	\left(
		U_{n-k}^\varepsilon
	\right)^p
\right] \frac{
	(2\tilde c \|v \|_\infty)^p
}{
	M^{p\left(
		\frac{1}{2}-\varepsilon
	\right)} K^\frac{p}{2}
}, \label{eqX5}
\end{equation}
where $\frac{1}{2}-\varepsilon > \frac{1}{2} - \frac{2}{q} \ge 0$ (see Lemma \ref{lmW}), since $q \ge 4$ in Assumption \ref{asExchange}. Moreover, from Lemma \ref{lmW}, there exists a constant $u^{\varepsilon,q} < \infty$ such that $\sup_{n\ge 0}  E\left[ \left( U_n^\varepsilon \right)^q \right] < u^{\varepsilon,q} < \infty$ for some $q \ge 4$. Therefore, for any $p \le q$ there exists $u^{\varepsilon,p}$ such that $\sup_{n\ge 0}  E\left[ \left( U_n^\varepsilon \right)^p \right] < u^{\varepsilon,p} < \infty$ and 
\begin{equation}
E\left[
	\left|
		(v, \pi_{n-k}^N) - (v,\bar \pi_{n-k}^N) 
	\right|^p
\right] \le  \frac{
	u^{\varepsilon,p} (2\tilde c\|v\|_\infty)^p
}{
	M^{p\left(
		\frac{1}{2}-\varepsilon
	\right)} K^\frac{p}{2}
}, \nonumber 
\end{equation}
which readily yields, for any $v \in B(\mX)$, 
\begin{equation}
\| (v, \pi_{n-k}^N) - (v,\bar \pi_{n-k}^N) \|_p \le \frac{
	2 u^\varepsilon \tilde c \|v\|_\infty
}{
	M^{\frac{1}{2} - \varepsilon} K^\frac{1}{2}
},
\label{eqX6}
\end{equation}
for any $\varepsilon \in \left( \frac{1+\epsilon}{q}, \frac{1}{2} \right)$, any $p \le q$ and where $u^\varepsilon = (u^{\varepsilon,p})^\frac{1}{p} < \infty$ is constant w.r.t. $M$, $K$ and $n$.

We handle the second term in \eqref{eqIneq2.25} by way of a similar argument. Let us define the $\sigma$-algebra $\mF_{n-k-1}^\infty = \bigcup_{M \ge 1} \mF_{n-k-1}^M$, where each term in the countable union is a generated $\sigma$-algebra, namely
$$
\mF_{n-k-1}^M = \sigma\left(x_{0:n-k-1}^{(m,j)}, \bar x_{1:n-k-1}^{(m,j)}; \quad 1 \le m \le M, \quad 1 \le j \le K \right)
$$ 
and recall 
that $\tilde \pi_n^N = \pi_n^N$ for every $n$ (see Remark \ref{rmMiscelanea}).  
For any $v \in B(\mX)$, we can decompose the integrals in the second term of \eqref{eqIneq2.25} as follows. On one hand, for $(v,\bar \Psi_{n-k}(\pi_{n-k-1}^N))$ we readily obtain
\ifdefined\ONECOLUMN
	\begin{eqnarray}
	(v,\Psi_{n-k}(\pi_{n-k-1}^N)) &=& (v,\Psi_{n-k}(\tilde \pi_{n-k-1}^N)) \nonumber\\
	&=& \frac{
		\sum_{m=1}^M \sum_{j=1}^{K} E[ v(\bar x_{n-k}^{(m,j)}) g_{n-k}^{y_{n-k}}(\bar x_{n-k}^{(m,j)}) | \mF_{n-k-1}^\infty]  \frac{
			W_{n-k-1}^{(m)*}
		}{
			K
		}
	}{
		\sum_{l=1}^M \sum_{i=1}^K E[ g_{n-k}^{y_{n-k}}(\bar x_{n-k}^{(l,i)}) | \mF_{n-k-1}^\infty]   \frac{
			W_{n-k-1}^{(l)*}
		}{
			K
		}
	} \nonumber\\
	&=& \frac{
		\left(
			E\left[
				vg_{n-k}^{y_{n-k}} | \mF_{n-k-1}^\infty
			\right], \xi_{n-k}^N
		\right)
	}{
		\left(
			E\left[
				g_{n-k}^{y_{n-k}} | \mF_{n-k-1}^\infty
			\right], \xi_{n-k}^N
		\right)
	}, \label{eqXXX-0}
	\end{eqnarray}
\else
	\begin{align}
		&
		(v,\Psi_{n-k}(\pi_{n-k-1}^N)) = (v,\Psi_{n-k}(\tilde \pi_{n-k-1}^N))
		= \nonumber\\ & =
		\frac{
			\sum_{m=1}^M \sum_{j=1}^{K} E[ v(\bar x_{n-k}^{(m,j)}) g_{n-k}^{y_{n-k}}(\bar x_{n-k}^{(m,j)}) | \mF_{n-k-1}^\infty]  \frac{
				W_{n-k-1}^{(m)*}
			}{
				K
			}
		}{
			\sum_{l=1}^M \sum_{i=1}^K E[ g_{n-k}^{y_{n-k}}(\bar x_{n-k}^{(l,i)}) | \mF_{n-k-1}^\infty]   \frac{
				W_{n-k-1}^{(l)*}
			}{
				K
			}
		}
		= \nonumber\\ & =
		\frac{
			\left(
				E\left[
					vg_{n-k}^{y_{n-k}} | \mF_{n-k-1}^\infty
				\right], \xi_{n-k}^N
			\right)
		}{
			\left(
				E\left[
					g_{n-k}^{y_{n-k}} | \mF_{n-k-1}^\infty
				\right], \xi_{n-k}^N
			\right)
		},
		\label{eqXXX-0}
	\end{align}
\fi
where $\xi_{n-k}^N = \sum_{m=1}^M \sum_{j=1}^K \frac{W_{n-k-1}^{(m)}}{K} \delta_{\bar x_{n-k}^{(m,j)}}$. On the other hand, the integral $(v,\bar \pi_{n-k}^N)$ can be similarly written as 
\begin{equation}
(v,\bar \pi_{n-k}^N) =
\frac{
	\sum_{m=1}^M\sum_{j=1}^K v(\bar x_{n-k}^{(m,j)}) g_{n-k}^{y_{n-k}}(\bar x_{n-k}^{(m,j)}) w_{n-k-1}^{(m,j)*}
}{
	\sum_{s=1}^M\sum_{i=1}^K g_{n-k}^{y_{n-k}}(\bar x_{n-k}^{(s,i)}) w_{n-k-1}^{(s,i)*}
}, \nonumber
\end{equation}
where the weights $w_{n-k-1}^{(m,j)*}$ are obtained after the exchange step. Since the map $\beta$ used for the exchange is deterministic and one-to-one, we can readily compute $(l,r) = \beta^{-1}(m,j)$ and, tracing back the particle exchange, we arrive at
$$
w_{n-k-1}^{(m,j)*} = \tilde w_{n-k-1}^{(l,r)*} = \frac{\bar W_{n-k-1}^{(l)*}}{K}.
$$
As a consequence, it is possible to rewrite the integral $(v,\bar \pi_{n-k}^N)$ as
\ifdefined\ONECOLUMN
	\begin{equation}
	(v,\bar \pi_{n-k}^N) =
	\frac{
		\sum_{l=1}^M\sum_{r=1}^K v(\bar x_{n-k}^{\beta(l,r)}) g_{n-k}^{y_{n-k}}(\bar x_{n-k}^{\beta(l,r)}) \bar W_{n-k-1}^{(l)*}
	}{
		\sum_{s=1}^M\sum_{i=1}^K g_{n-k}^{y_{n-k}}(\bar x_{n-k}^{\beta(s,i)}) \bar W_{n-k-1}^{(s)*}
	}
	= \frac{
		(vg_{n-k}^{y_{n-k}}, \xi_{n-k}^N)
	}{
		(g_{n-k}^{y_{n-k}}, \xi_{n-k}^N)
	}. \label{eqXXX-1}
	\end{equation}
\else
	\begin{align}
		(v,\bar \pi_{n-k}^N)
		& =
		\frac{
			\sum_{l=1}^M\sum_{r=1}^K v(\bar x_{n-k}^{\beta(l,r)}) g_{n-k}^{y_{n-k}}(\bar x_{n-k}^{\beta(l,r)}) \bar W_{n-k-1}^{(l)*}
		}{
			\sum_{s=1}^M\sum_{i=1}^K g_{n-k}^{y_{n-k}}(\bar x_{n-k}^{\beta(s,i)}) \bar W_{n-k-1}^{(s)*}
		}
		\nonumber \\ & =
		\frac{
			(vg_{n-k}^{y_{n-k}}, \xi_{n-k}^N)
		}{
			(g_{n-k}^{y_{n-k}}, \xi_{n-k}^N)
		}.
		\label{eqXXX-1}
	\end{align}
\fi
Combining \eqref{eqXXX-0} and \eqref{eqXXX-1}, and after some straightforward algebraic manipulations, the difference $(v,\bar \pi_{n-k}^N) - (v,\Psi_{n-k}(\pi_{n-k-1}^N))$ can be rewritten as
\ifdefined\ONECOLUMN
	\begin{eqnarray}
	(v,\bar \pi_{n-k}^N) - (v,\Psi_{n-k}(\pi_{n-k-1}^N)) &=& \nonumber\\
	\frac{1}{(g_{n-k}^{y_{n-k}},\xi_{n-k}^{N})} \times \left(
		(vg_{n-k}^{y_{n-k}},\xi_{n-k}^N) - \left(
			E\left[
				vg_{n-k}^{y_{n-k}}|\mF_{n-k-1}^\infty
			\right]
		, \xi_{n-k}^N
	\right) \right) && \nonumber \\
	+ \frac{
		\| v \|_\infty 
	}{
		(g_{n-k}^{y_{n-k}},\xi_{n-k}^{N})
	} \times \left(
		(g_{n-k}^{y_{n-k}},\xi_{n-k}^N) - \left(
			E\left[
				g_{n-k}^{y_{n-k}}|\mF_{n-k-1}^\infty
			\right], \xi_{n-k}^N
		\right)
	\right). &&\nonumber
	\end{eqnarray}
\else
	\begin{align*}
		&
		(v,\bar \pi_{n-k}^N) - (v,\Psi_{n-k}(\pi_{n-k-1}^N))
		=
		\frac{1}{(g_{n-k}^{y_{n-k}},\xi_{n-k}^{N})}
		\times \\ & \times
		\left(
			(vg_{n-k}^{y_{n-k}},\xi_{n-k}^N) - \left(
				E\left[
					vg_{n-k}^{y_{n-k}}|\mF_{n-k-1}^\infty
				\right]
			, \xi_{n-k}^N
		\right) \right)
		+ \\ & +
		\frac{
			\| v \|_\infty 
		}{
			(g_{n-k}^{y_{n-k}},\xi_{n-k}^{N})
		} \times \left(
			(g_{n-k}^{y_{n-k}},\xi_{n-k}^N)
		- \right. \\ & - \left.
			\left(
				E\left[
					g_{n-k}^{y_{n-k}}|\mF_{n-k-1}^\infty
				\right], \xi_{n-k}^N
			\right)
		\right)
		.
	\end{align*}
\fi
Resorting to Minkowski's inequality, Assumption \ref{asBounded} and the fact that all integrals are computed w.r.t. the same measure, $\xi_{n-k}^N$, the equality above easily yields the bound
\ifdefined\ONECOLUMN
	\begin{eqnarray}
	\left\| 
		(v,\bar \pi_{n-k}^N) - (v,\Psi_{n-k}(\pi_{n-k-1}^N))
	\right\|_p &\le& a \left\|
		\left(
			vg_{n-k}^{y_{n-k}} - E\left[
				vg_{n-k}^{y_{n-k}}|\mF_{n-k-1}^\infty
			\right] , \xi_{n-k}^N
		\right)
	\right\|_p \nonumber \\
	&+& a \| v \|_\infty \left\|
		\left(
			g_{n-k}^{y_{n-k}} - E\left[
				g_{n-k}^{y_{n-k}}|\mF_{n-k-1}^\infty
			\right], \xi_{n-k}^N
		\right)
	\right\|_p. \nonumber
	\end{eqnarray}
\else
	\begin{align*}
		&
		\left\| 
			(v,\bar \pi_{n-k}^N) - (v,\Psi_{n-k}(\pi_{n-k-1}^N))
		\right\|_p
		\le \\ & \le
		a \left\|
			\left(
				vg_{n-k}^{y_{n-k}} - E\left[
					vg_{n-k}^{y_{n-k}}|\mF_{n-k-1}^\infty
				\right] , \xi_{n-k}^N
			\right)
		\right\|_p
		+ \\ & +
		a \| v \|_\infty \times \left\|
			\left(
				g_{n-k}^{y_{n-k}} - E\left[
					g_{n-k}^{y_{n-k}}|\mF_{n-k-1}^\infty
				\right], \xi_{n-k}^N
			\right)
		\right\|_p
	\end{align*}
\fi
However, the integral $\left(
	vg_{n-k}^{y_{n-k}} - E\left[
		vg_{n-k}^{y_{n-k}}|\mF_{n-k-1}^\infty
	\right] , \xi_{n-k}^N
\right)$ with $v \in B(\mX)$ reduces to
\ifdefined\ONECOLUMN
	\begin{equation}
	\left(
		vg_{n-k}^{y_{n-k}} - E\left[
			vg_{n-k}^{y_{n-k}}|\mF_{n-k-1}^\infty
		\right] , \xi_{n-k}^N
	\right) = \sum_{m=1}^M \sum_{j=1}^K \frac{\bar W_{n-k-1}^{(m)}}{K} \breve Z_{n-k}^{(m,j)},
	\label{eqX7}
	\end{equation}
\else
	\begin{align}
		&
		\left(
			vg_{n-k}^{y_{n-k}} - E\left[
				vg_{n-k}^{y_{n-k}}|\mF_{n-k-1}^\infty
			\right] , \xi_{n-k}^N
		\right)
		= \nonumber\\ & =
		\sum_{m=1}^M \sum_{j=1}^K \frac{\bar W_{n-k-1}^{(m)}}{K} \breve Z_{n-k}^{(m,j)},
		\label{eqX7}
	\end{align}
\fi
where, for all $m \in \{1, ..., M\}$ and $j \in \{1, ..., K\}$,
\ifdefined\ONECOLUMN
	$$
	\breve Z_{n-k}^{(m,j)} = v(\bar x_{n-k}^{(m,j)}) g_{n-k}^{y_{n-k}}(\bar x_{n-k}^{(m,j)}) - E[ v(\bar x_{n-k}^{(m,j)}) g_{n-k}^{y_{n-k}}(\bar x_{n-k}^{(m,j)}) | \mF_{n-k-1}^\infty ],
	$$
\else
	\begin{align*}
		\breve Z_{n-k}^{(m,j)}
		= {} &
		v(\bar x_{n-k}^{(m,j)}) g_{n-k}^{y_{n-k}}(\bar x_{n-k}^{(m,j)})
		- \\ & -
		E[ v(\bar x_{n-k}^{(m,j)}) g_{n-k}^{y_{n-k}}(\bar x_{n-k}^{(m,j)}) | \mF_{n-k-1}^\infty ]
	\end{align*}
\fi
are conditionally independent r.v.'s, with zero mean and bounded as $| \breve Z_{n-k-1}^{(m,j)} | \le 2 a \| v \|_\infty$ (recall that $\| g_n^{y_n} \|_\infty < a$ for every $n$, from Assumption \ref{asBounded}). Therefore, using exactly the same argument that led us from Eq. \eqref{eqEq2.5} to the inequality \eqref{eqX6} (involving the use of Assumption \ref{asExchange} to upper-bound the aggregate weights) now we arrive at
\begin{equation}
\left\| 
	(v,\bar \pi_{n-k}^N) - (v,\Psi_{n-k}(\pi_{n-k-1}^N))
\right\|_p \le \frac{
	4 a^2 u^\varepsilon \tilde c \| v \|_\infty
}{
	M^{\frac{1}{2} - \varepsilon} K^\frac{1}{2}
},
\label{eqX8}
\end{equation}
for any $\frac{1+\epsilon}{q} < \varepsilon < \frac{1}{2}$ and $p \le q$, where $\tilde c<\infty$, $a < \infty$ and $u^\varepsilon<\infty$ are constants w.r.t. $M$, $K$ and $n$.

Next, we substitute backwards to complete the proof. First, we insert \eqref{eqX6} and \eqref{eqX8} into the triangle inequality \eqref{eqIneq2.25}, to obtain
\begin{equation}
\| (v,\pi_{n-k}^N) - ( v,\Psi_{n-k}(\pi_{n-k-1}^N) ) \|_p \le \frac{
	8a^2 u^\varepsilon \tilde c \| v \|_\infty
}{
	M^{\frac{1}{2}-\varepsilon} K^\frac{1}{2}
},
\label{eqX9}
\end{equation}
which yields a bound for error terms of the form in \eqref{eqBoudForE}, by simply taking $v=\Gamma_{n|n-k}(h)$. Using this bound in \eqref{eqIneq2} we arrive at inequality \eqref{eqX10} in the statement of Lemma \ref{lmPsi}. $\qed$

\begin{Teorema} \label{thUniform1}
Let $K<\infty$ be fixed. If Assumptions \ref{asBounded}, \ref{asStable} and \ref{asExchange} hold, then the approximate measures computed via the DRNA algorithm converge uniformly over time in $L_p$. To be specific,
\begin{equation}
\lim_{M \rw \infty} \sup_{n\ge 0} \left\|
	(h,\pi_n^{MK}) - (h,\pi_n)
\right\|_p = 0
\label{eqLimUnif}
\end{equation}
for any $h \in B(\mX)$ and every $1 \le p \le q$, where $q \ge 4$ is given by Assumption \ref{asExchange}.  

\end{Teorema}

\noindent {\bf Proof.} Again, we write $\pi_n^N=\pi_n^{MK}$ for conciseness. The proof follows the same kind of argument as in \cite{DelMoral01c}. Let us choose an arbitrary integer $T > 1$ and look into the error terms for $n \le T$ and $n > T$ separately. For $n \le T$, the difference $(h,\pi_n^N)-(h,\pi_n)$ can be easily decomposed as 
\ifdefined\ONECOLUMN
	\begin{eqnarray}
		(h,\pi_n^N)-(h,\pi_n)
		&=& \left(
			\sum_{k=0}^{n-1} (h, \Psi_{n|n-k}(\pi_{n-k}^N)) - (h,\Psi_{n|n-k-1}(\pi_{n-k-1}^N))
		\right)\nonumber \\
		&& + (h,\Psi_{n|0}(\pi_0^N)) - (h,\Psi_{n|0}(\pi_0)) \nonumber
	\end{eqnarray}
\else
	\begin{eqnarray}
		(h,\pi_n^N)-(h,\pi_n)
		&=& \nonumber\\
		\left(
			\sum_{k=0}^{n-1} (h, \Psi_{n|n-k}(\pi_{n-k}^N)) - (h,\Psi_{n|n-k-1}(\pi_{n-k-1}^N))
		\right) && \nonumber \\
		+ (h,\Psi_{n|0}(\pi_0^N)) - (h,\Psi_{n|0}(\pi_0)) && \nonumber
	\end{eqnarray}
\fi
hence we readily find an upper bound for the approximation error in $L_p$ with a similar structure, namely
\ifdefined\ONECOLUMN
	\begin{eqnarray}
	\| (h,\pi_n^N)-(h,\pi_n) \|_p &\le& 
	\sum_{k=0}^{n-1} \|
		(h, \Psi_{n|n-k}(\pi_{n-k}^N)) - (h,\Psi_{n|n-k-1}(\pi_{n-k-1}^N))
	\|_p \nonumber \\
	&& + \|
		(h,\Psi_{n|0}(\pi_0^N)) - (h,\Psi_{n|0}(\pi_0))
	\|_p. \label{eqIneq0}
	\end{eqnarray}
\else
	\begin{align}
		\| (h,\pi_n^N)-(h,\pi_n) \|_p
		\le {} &
		\sum_{k=0}^{n-1} \|
			(h, \Psi_{n|n-k}(\pi_{n-k}^N))
		- \nonumber\\ & -
		(h,\Psi_{n|n-k-1}(\pi_{n-k-1}^N))
		\|_p
		+ \nonumber\\ & +
		\|
			(h,\Psi_{n|0}(\pi_0^N)) - (h,\Psi_{n|0}(\pi_0))
		\|_p
		.
		\label{eqIneq0}
	\end{align}
\fi

For the second term on the right hand side (r.h.s.) of \eqref{eqIneq0} we have
\ifdefined\ONECOLUMN
	\begin{eqnarray}
	\| (h,\Psi_{n|0}(\pi_0^N)) - (h,\Psi_{n|0}(\pi_0)) \|_p &=& 
	\left\|
			\frac{
				(\Gamma_{n|0}(h), \pi_0^N)
			}{
				(\Gamma_{n|0}(1), \pi_0^N)
			} - \frac{
				(\Gamma_{n|0}(h), \pi_0)
			}{
				(\Gamma_{n|0}(1), \pi_0)
			} 
	\right\|_p \label{eqEq1} \\
	&\le& a^n \| (\Gamma_{n|0}(h), \pi_0^N) - (\Gamma_{n|0}(h), \pi_0) \|_p \label{eqIq2} \\
	&&+  a^n \| h \|_\infty \| (\Gamma_{n|0}(1), \pi_0^N) - (\Gamma_{n|0}(1), \pi_0) \|_p, \nonumber 
	\end{eqnarray}
\else
	\begin{align}
		&
		\| (h,\Psi_{n|0}(\pi_0^N)) - (h,\Psi_{n|0}(\pi_0)) \|_p
		= \nonumber\\ & =
		\left\|
				\frac{
					(\Gamma_{n|0}(h), \pi_0^N)
				}{
					(\Gamma_{n|0}(1), \pi_0^N)
				} - \frac{
					(\Gamma_{n|0}(h), \pi_0)
				}{
					(\Gamma_{n|0}(1), \pi_0)
				} 
		\right\|_p \label{eqEq1}
		\\ & \le
		a^n \| (\Gamma_{n|0}(h), \pi_0^N) - (\Gamma_{n|0}(h), \pi_0) \|_p 
		+ \nonumber \\ & +
		a^n \| h \|_\infty \| (\Gamma_{n|0}(1), \pi_0^N) - (\Gamma_{n|0}(1), \pi_0) \|_p
		,
		\label{eqIq2}
	\end{align}
\fi
where the equality \eqref{eqEq1} follows from Eq. \eqref{eqPsiGamma} while \eqref{eqIq2} is a consequence of the inequality \eqref{eqPreliminaries} together with Assumption \ref{asBounded}. Since it is straightforward to show that $\Gamma_{n|0}(h)$ and $\Gamma_{n|0}(1)$ are bounded, namely 
$$
\| \Gamma_{n|0}(h) \|_\infty \le a^n \| h \|_\infty \quad (\mbox{hence} \quad \| \Gamma_{n|0}(1) \|_\infty \le a^n),
$$
and $\pi_0^N$ is built with independent and identically distributed (i.i.d.) samples from $\pi_0$, we readily obtain the usual Monte Carlo bound for the approximation error of $(h,\Psi_{n|0}(\pi_0^N))$, i.e.,
\begin{equation}
\| (h,\Psi_{n|0}(\pi_0^N)) - (h,\Psi_{n|0}(\pi_0)) \|_p \le \frac{
	C \| h \|_\infty a^n
}{
	\sqrt{MK}
} \le  \frac{
	C \| h \|_\infty a^T
}{
	\sqrt{MK}
}
\label{eqIneq1}
\end{equation}
where $C$ is a constant independent of $M$, $K$, $p$ and $n$, and the second inequality holds because we are looking at the case $n \le T$.

The terms in the summation of \eqref{eqIneq0} can be upper-bounded using Lemma \ref{lmPsi} (note that the assumptions of Lemma \ref{lmPsi} are a subset of the assumptions in Theorem \ref{thUniform1}). Indeed, combining the inequality \eqref{eqX10} and the bound in \eqref{eqIneq1} into the original inequality \eqref{eqIneq0} yields
\begin{align}
	\| (h,\pi_n^N)-(h,\pi_n) \|_p
	& \le
	\frac{
		8 n a^{n+1} u^\varepsilon \tilde c \| h \|_\infty
	}{
		M^{\frac{1}{2}-\varepsilon} K^\frac{1}{2}
	} + \frac{
		C \| h \|_\infty a^T
	}{
		\sqrt{MK}
	} \label{eqX10.5}
	\\ & \le
	\frac{
		8 T a^{T+1} \tilde C \| h \|_\infty
	}{
		M^{\frac{1}{2}-\varepsilon} K^\frac{1}{2}
	}
	,
	\label{eqX11}
\end{align}
where $\tilde C = \max\{u^\varepsilon \tilde c,C\}$ and we have taken into account that there are at most $T$ terms in the summation of \eqref{eqIneq0} in order to obtain the second inequality.

Similar to \eqref{eqIneq0}, for $n > T$ we have
\ifdefined\ONECOLUMN
	\begin{eqnarray}
	\| (h,\pi_n^N) - (h,\pi_n) \|_p &\le&
	\sum_{k=0}^{T-1} \| (h,\Psi_{n|n-k}(\pi_{n-k}^N)) - (h,\Psi_{n|n-k-1}(\pi_{n-k-1}^N)) \|_p \nonumber \\
	&& + \| (h,\Psi_{n|n-T}(\pi_{n-T}^N)) - (h,\Psi_{t|n-T}(\pi_{n-T})) \|_p. 
	\label{eqX12}
	\end{eqnarray}
\else
	\begin{align}
		{} &
		\| (h,\pi_n^N) - (h,\pi_n) \|_p
		\le \nonumber \\ {} &
		\sum_{k=0}^{T-1} \| (h,\Psi_{n|n-k}(\pi_{n-k}^N)) - (h,\Psi_{n|n-k-1}(\pi_{n-k-1}^N)) \|_p
		+ \nonumber \\ & +
		\| (h,\Psi_{n|n-T}(\pi_{n-T}^N)) - (h,\Psi_{t|n-T}(\pi_{n-T})) \|_p. 
		\label{eqX12}
	\end{align}
\fi
The remainder term on the rhs of \eqref{eqX12} can be directly bounded by way of Assumption \ref{asStable}, namely
\begin{equation}
\| (h,\Psi_{n|n-T}(\pi_{n-T}^N)) - (h,\Psi_{t|n-T}(\pi_{n-T})) \|_p \le \mE(h,T),
\label{eqX13}
\end{equation}
where $\lim_{T\rw\infty} \mE(h,T) = 0$ for every $h \in B(\mX)$. The summation on the rhs of \eqref{eqX12}, on the other hand, has exactly the same structure as the summation in \eqref{eqIneq0}, hence the bound in \eqref{eqX11} is still valid here and we can combine it with \eqref{eqX13} and \eqref{eqX12} to arrive at\footnote{It is, indeed, important to realize at this point that the factor $n$ in the numerator $8 n a^{n+1} u^\varepsilon \tilde c \| h \|_\infty$ of the inequality \eqref{eqX10.5} arises exclusively from the number of terms in the summation of \eqref{eqIneq0}, which is at most $T$, and not because of an actual dependence on time. Therefore, exactly the same argument is valid for the summation of \eqref{eqX12}, even if $n > T$.}
\begin{equation}
\| (h,\pi_n^N) - (h,\pi_n) \|_p \le \frac{
	8 T a^{T+1} \tilde C \| h \|_\infty
}{
	M^{\frac{1}{2}-\varepsilon} K^\frac{1}{2}
} + \mE(h,T)
\label{eqX14}
\end{equation}
for $n>T$. Since the bound above is independent of $n$, and valid for arbitrary $T$, taking together \eqref{eqX11} and \eqref{eqX14} yields
\begin{equation}
\sup_{n \ge 0} \| (h,\pi_n^N) - (h,\pi_n) \|_p \le \frac{
	8 T a^{T+1} \tilde C \| h \|_\infty
}{
	M^{\frac{1}{2}-\varepsilon} K^\frac{1}{2}
} + \mE(h,T).
\label{eqX15}
\end{equation}

Finally, for $M$ sufficiently large, if we choose $T=T_M$ and
\begin{equation}
T_M= \left\lfloor
	\frac{
		(\frac{1}{2}-\varepsilon-\gamma)\log(M) - \log(a) - \log( 8 \tilde C \| h \|_\infty )
	}{
		1 + \log a
	}
\right\rfloor
\label{eqDefT_M}
\end{equation}
for any $\gamma \in \left( 0,\frac{1}{2}-\varepsilon \right)$, then 
\begin{equation}
\frac{
	8 T a^{T+1} \tilde C \| h \|_\infty
}{
	M^{\frac{1}{2}-\varepsilon} K^\frac{1}{2}
} \le \frac{
	1
}{
	M^\gamma K^\frac{1}{2}
}.
\label{eqPrelim_Rate}
\end{equation}
Since $\lim_{M\rw\infty} T_M = \infty$ it follows that $\lim_{M\rw\infty} \mE(h,T_M) = 0$ and, therefore, for $N=MK$, 
$
\lim_{M\rw\infty} \sup_{n\ge 0} \|(h,\pi_n^N)-(h,\pi_n)\|_p = 0. \quad \qed
$


\subsection{Convergence rates}

Theorem \ref{thUniform1} provides a theoretical guarantee for the convergence of Algorithm \ref{alDPF} when the number of PEs increases (even for a fixed number, $K$, of particles per PE). However, it does not provide a convergence rate and, as a consequence, it is not possible to compare its performance with conventional (centralized) PFs, for which convergence rates are well known (see, e.g., \cite{DelMoral01c,DelMoral04,Bain08}). Following an approach similar to \cite{DelMoral01c} for the standard PF, we show in this section that it is possible to obtain an explicit convergence rate for Algorithm \ref{alDPF} when the optimal filter is stable with a known rate itself. In particular, we adopt the following assumption, which entails the exponential stability of the optimal filter.

\begin{Hipotesis} \label{asExpStable}
For any $\alpha, \eta \in \mP(\mX)$ and every $h \in B(\mX)$, there exist constants $T_0 < \infty$ and $\nu > 0$ such that
$$
\sup_{n \ge 0} \left|
	(h,\Psi_{n+T|n}(\alpha)) - (h,\Psi_{n+T|n}(\eta))
\right| < \exp\left\{
	-\nu T
\right\} \quad \mbox{for every $T > T_0$.}
$$
\end{Hipotesis}

See \cite{DelMoral01c,DelMoral04,Papavasiliou06} for a discussion of sufficient conditions for exponential stability. Using Assumption \ref{asExpStable} we can strengthen Theorem \ref{thUniform1} in order to obtain the following result.

\begin{Teorema} \label{thRate}
Let $K<\infty$ be fixed. If Assumptions \ref{asBounded}--\ref{asExpStable} hold, then for any $h \in B(\mX)$ and $p \le q$
\begin{equation}
\sup_{n\ge 0} \left\|
	(h,\pi_n^{MK}) - (h,\pi_n)
\right\|_p \le \frac{
	C
}{
	M^\zeta K^\frac{1}{2}
}
\label{eqThRate}
\end{equation}
for some $C<\infty$ and $\zeta > 0$ independent of $M$ and $K$. In particular, 
\begin{equation}
\zeta = \min\left\{
	1, \frac{\nu}{1+\log(a)}
\right\} \Upsilon,
\label{eqZeta}
\end{equation}
for any $\Upsilon  \in \left( \frac{1+\epsilon}{q}, \frac{1}{2} - \frac{1+\epsilon}{q} \right)$, with $\epsilon \in (0,1)$ and $q \ge 4$ given by Assumption \ref{asExchange}. 
\end{Teorema}

\begin{Nota}
If Assumption \ref{asExchange} holds for arbitrarily large $q$ then the inequality \eqref{eqThRate} holds for $\zeta$ arbitrarily close to $\frac{1}{2} \tilde \nu$, where the coefficient $\tilde \nu = \min\left\{ 1, \frac{\nu}{1+\log(a)} \right\}$ depends on the state-space system. In this case, Algorithm \ref{alDPF} matches the convergence rate obtained for the standard PF with the same kind of analysis \cite{DelMoral04}. If  Assumption \ref{asExchange} holds only for some relatively small $q\ge 4$, then there is an actual loss in the convergence rate of Algorithm \ref{alDPF} compared to a centralized PF.
\end{Nota}

\noindent {\bf Proof of Theorem \ref{thRate}.} Again, let us write $\pi_n^{MK} = \pi_n^N$ for conciseness. We recall Eq. \eqref{eqX15}, reproduced below for convenience,
\begin{equation}
\sup_{n \ge 0} \| (h,\pi_n^N) - (h,\pi_n) \|_p \le \frac{
	8 T a^{T+1} \tilde C \| h \|_\infty
}{
	M^{\frac{1}{2}-\varepsilon} K^\frac{1}{2}
} + \mE(h,T),
\label{eqX15bis}
\end{equation}
where $\varepsilon \in \left( \frac{\epsilon+1}{q}, \frac{1}{2} \right)$ and $\tilde C<\infty$ is constant w.r.t. $M$ and $K$. Since $T<\infty$ is arbitrary, we can $T=T_M$ like in \eqref{eqDefT_M} 
which allows to upper-bound the first term on the right hand side of \eqref{eqX15bis} using the inequality \eqref{eqPrelim_Rate}.
As for the second term on the rhs of \eqref{eqX15bis}, Assumption \ref{asExpStable} yields (for large $M$, so that $T_M \ge T_0$), 
\begin{equation}
\mE(h,T_M) \le \exp\left\{
	- \nu T_M
\right\} 
\le \frac{
	C
}{
	M^{
		\frac{
			\nu	
		}{
			1 + \log(a)
		} \left(
			\frac{1}{2} - \gamma - \varepsilon
		\right)
	}
}, \label{eqChunga2}
\end{equation}
where 
$
C = - \frac{
	\log(a) + \log( 8 \tilde C \| h \|_\infty )
}{
	1 + \log(a)
}.
$

Combining the inequalities \eqref{eqPrelim_Rate} and \eqref{eqChunga2}
into \eqref{eqX15bis} yields the inequality \eqref{eqThRate} in Theorem \ref{thRate}, with $\Upsilon = \frac{1}{2} - \varepsilon - \gamma \in \left( \frac{1+\epsilon}{q}, \frac{1}{2}-\frac{1+\epsilon}{q} \right)$ in \eqref{eqZeta}. $\qed$



\subsection{Discussion}

In this section we have proved that Algorithm \ref{alDPF}, based on the DRNA scheme of \cite{Bolic05}, converges asymptotically with the number of PEs, $M$, and uniformly over time. This is, to our best knowledge, the first rigorous proof of convergence for this type of PF, which has been used extensively in the literature \cite{Miguez07b,Lozano09,Ahmed10,Hendeby10,Balasingam11,Read14}. Note that classical analyses, such as in \cite{Crisan01,Crisan02,DelMoral01c,DelMoral04,Bain08}, do not hold for this algorithm because they do not take into account the distinct aggregate weights of the subsets of particles assigned to the PEs.

Uniform convergence over time is relevant from a practical point of view. It implies that the DPF can run for an indefinitely long period of time, since the computational load (i.e., the number of particles) needed to guarantee a certain error bound (namely, the rhs of inequality \eqref{eqThRate}) is independent of the time index $n$. This should be compared with the classical convergence analyses in \cite{Bain08} or \cite{Miguez13b}, which are based on induction arguments and yield error bounds of the form $C_n/\sqrt{N}$, where $N$ is the number of particles and $C_n$ is a constant independent of $N$. It is simple to show \cite{Bain08} that $\lim_{n\rw\infty} C_n = \infty$ and, therefore, the error bound diverges with time, i.e., $\lim_{n\rw\infty} \frac{C_n}{N} = \infty$ for any fixed $N$. Based on this type of analysis, the PF can only be guaranteed to work for a finite period of time. The advantage of uniform convergence, where the error bound is independent of time, comes at the expense of additional assumptions (Assumptions \ref{asBounded}--\ref{asExpStable} in our case) which are not needed for the proofs of \cite{Bain08} or \cite{Miguez13b}. 

Assumptions \ref{asBounded}, \ref{asStable} and \ref{asExpStable} are related to the stability of the optimal filter for the state-space model of interest. They refer to properties of the model, which may hold or not independently of the filtering algorithm we use. In practice, it is usually easy to show that they hold for models in which the state space $\mX$ is compact. This kind of assumptions is common in the literature \cite{DelMoral01c,LeGland04,Papavasiliou05}. 

Assumption \ref{asExchange} is similar to the regularity conditions imposed on the weights in \cite{Douc07,Cornebise08}. We investigate its validity numerically, by way of computer simulations, in the example of Section \ref{sSimulations}. Intuitively, it implies that the aggregate weights of the PEs remain ``balanced'', i.e., no PE is expected to accumulate all the weight --a situation that would lead to degeneracy of the distributed scheme, which would be reduced to a centralized PF with only $K$ particles. Let us also point out that it is possible to monitor the aggregate weights online and possibly schedule additional exchange steps in order to guarantee, e.g., that $W_n^{(m)^4} \le cst / M^{4-\epsilon}$ for every $m$. The latter inequality is much stronger than Assumption \ref{asExchange}, yet has the advantage of being verifiable in practice. 

Theorem \ref{thRate} provides an explicit rate for the uniform convergence of Algorithm \ref{alDPF}. This is relevant because it can be argued that Theorem \ref{thUniform1} alone does not guarantee a ``practically acceptable'' performance. In particular, even if \eqref{eqLimUnif} holds, convergence may still be so slow that the filter cannot be used for any practical purpose. We foresee that it may be possible to improve the error rate in Theorem \ref{thRate} by using techniques borrowed from \cite{Whiteley13}, which relies on slightly stronger assumptions on the state space model {\em and} the algorithm. 

\section{Computer simulations} \label{sSimulations}

%
\subsection{State space model}

We have carried out computer simulations for a problem consisting in the tracking of a target that moves over a 2-dimensional rectangular region, using a WSN consisting of $J$ nodes that produce binary (0 or 1) outputs, depending on the distance between the target and the node. In the sequel we describe the state space model for this problem.

The system state at time $n$ is denoted $x_n=[r_n,v_n]^\top \in \Real^4$, where $r_n \in \Real^2$ is the target position and $v_n\in\Real^2$ is the target velocity. The prior distribution has the form $\tau_0(dx_0) = \mU(r_0; \mR)dr_0 \times \mN(v_0; 0, \sigma_{r,0}^2 I_2)dv_0$, where $\mR = [-20,20] \times [-10,10]$ is the rectangular region of interest, $\mU(r_0; \mR)$ is the uniform pdf over $\mR$ of the initial position, $r_0$, and  $\mN(v_0; 0, \sigma_{v,0}^2 I_2)$ is the Gaussian pdf of the initial velocity, $v_0$, which has zero mean and covariance function $\sigma_{v,0}^2 I_2$, with $I_2$ the $2\times 2$ identity matrix. The variance parameter is $\sigma_{v,0} = 5\times 10^{-2}$ for all simulations.

\begin{figure}[htpb]
\begin{center}
	\includegraphics[width=0.8\linewidth]{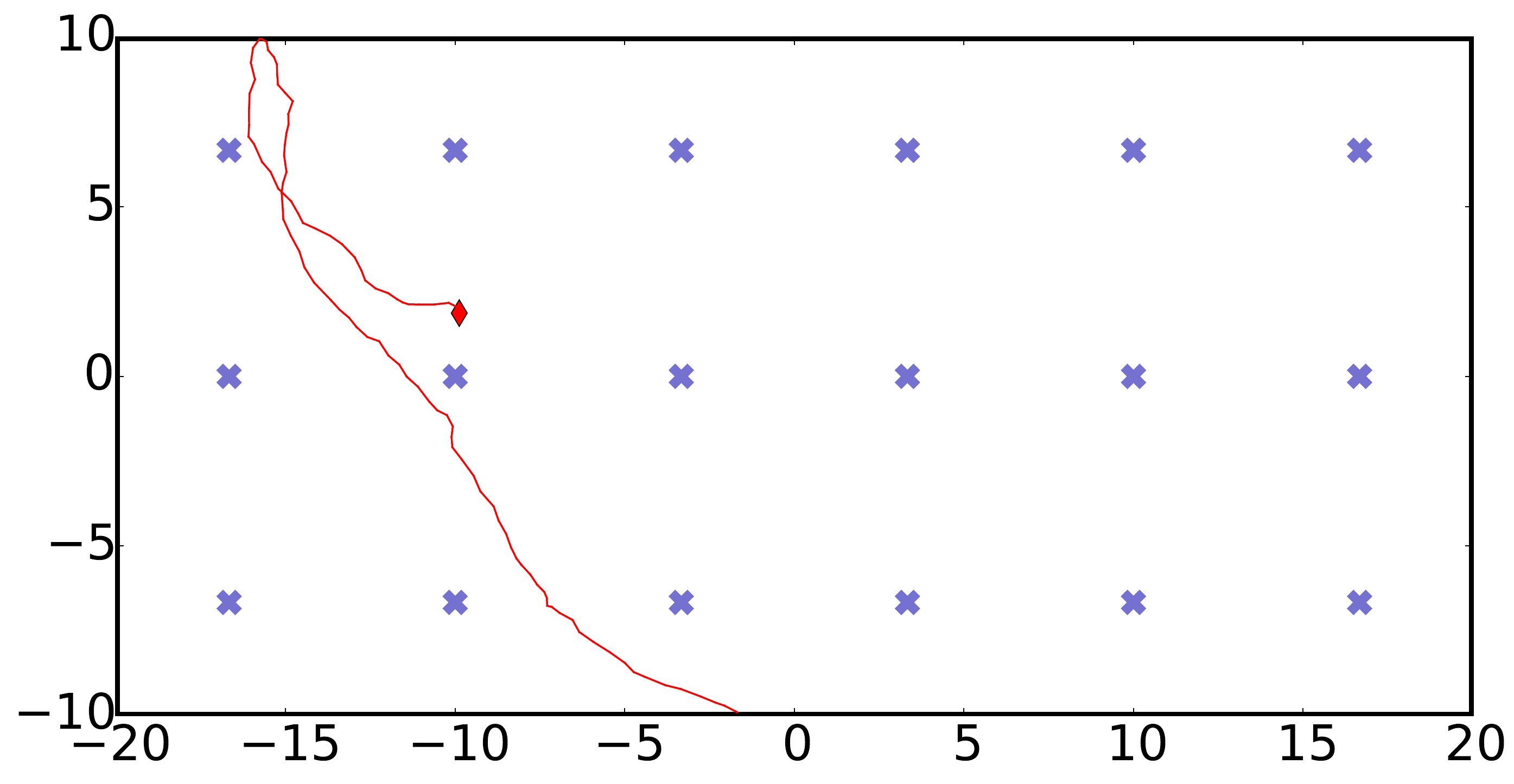}
\end{center}
\caption{Sample trajectory of the target during the first $100$ time instants. The crosses mark the positions of the sensors, whereas the diamond indicates the starting point. Vertical and horizontal axes in meters.}
\label{fTrajectory}
\end{figure}

In order to apply either Algorithm \ref{alDPF} or a centralized standard PF we need to describe how to produce random samples from the transition kernel $\tau_n(dx_n|x_{n-1})$. Given the state at time $n-1$, let us introduce the auxiliary r.v. 
\begin{equation}
\tilde x_n = \left[
	\begin{array}{c}
	\tilde r_n\\
	\tilde v_n\\
	\end{array}
\right] = \left[
	\begin{array}{cc}
	I_2 &\kappa I_2\\
	0 &I_2
	\end{array}
\right] x_{n-1} + \eta_n,
\nonumber
\end{equation}
where $\kappa$ is the duration of the discrete time steps in the model (i.e., the continuous time elapsed between two consecutive realizations of the system state), and $\eta_n$ is a sequence of i.i.d. Gaussian r.v.'s with pdf $\mN(\eta_n; 0, C_\eta)$, where the covariance matrix  has the form 
$C_\eta = \left[
	\begin{array}{cc}
	(\kappa^2\sigma_v^2 + \sigma_r^2)I_2 &0\\
	0 &\sigma_v^2 I_2\\
	\end{array}
\right]$ and the parameters $\sigma_v^2$ and $\sigma_r^2$ represent the variance of any unknown (random) acceleration effects and other direct random perturbations of the target position, respectively. We set $\kappa=1$ and $\sigma_r^2=\sigma_v^2=10^{-2}$ for the simulations. We also introduce a sequence of i.i.d. Gaussian r.v.'s $u_n$, $n \ge 1$, with the same distribution as the initial velocity, i.e., $\mN(u_n; 0, \sigma_{v,0}I_2)$. Then, the state $x_n$ conditional on $x_{n-1}$ can be generated as
\begin{equation}
x_n = \left\{
	\begin{array}{cl}
	\tilde x_n, &\mbox{if $\tilde r_n \in \mR$},\\
	\left[r_{n-1}, u_n\right]^\top, &\mbox{if $\tilde r_n \notin \mR$},\\
	\end{array}
\right..
\nonumber
\end{equation} 
A sample realization of a target trajectory, during 100 discrete time steps, according to the described model can be seen in Figure \ref{fTrajectory}.

The WSN consists of $J=18$ binary sensors. The $j$-th sensor position is denoted $s_j \in \mR$ and its output is $y_n(j) \in \{0,1\}$, hence the complete observation vector at time $n$ is $y_n = [y_n(1), ..., y_n(J)]^\top$. The sensors measure whether the target appears to lie within a threshold distance $\mu=7$ m of the sensor position, but the output is random. To be specific, the output $y_n(j)$ conditional on $\| r_n - s_j \| \le \mu$ is a Bernoulli r.v. with parameter $p_1$, whereas $y_n(j)$ conditional on  $\| r_n - s_j \| > \mu$ is Bernoulli with parameter $\bar p_1$. We refer to $p_1$ as the detection probability, and set $p_1=0.9$ for the simulations, while we set $\bar p_1=10^{-2}$ and we refer to it as the false alarm probability.

The probability mass of the observations can be written using the general notation in Section \ref{ssStateSpaceModel} as $g_n(y_n|x_n) = \prod_{j=1}^J g_n(y_n(j)|x_n)$, where
$$
g_n(y_n(j)|x_n) = \left\{
	\begin{array}{cl}
	p_1 y_n(j) + (1-p_1)(1-y_n(j)), &\mbox{if $\|r_n-s_j\|\le\mu$}\\
	\bar p_1 y_n(j) + (1-\bar p_1)(1-y_n(j)), &\mbox{if $\|r_n-s_j\|>\mu$}\\
	\end{array}
\right.,
$$
which allows to compute any necessary importance weights.

%
\subsection{Numerical results} \label{ssNumerical}

We first assess the validity of Assumption \ref{asExchange}, which is key in the analysis of Section \ref{sAnalysis}, and then compare the performance of the DPF described by Algorithm \ref{alDPF} with a standard (centralized) PF in terms of the absolute error of the position estimates. Note that Algorithm \ref{alDPF} actually reduces to a standard (or {\em bootstrap}) PF if we simply set $M=1$ and, therefore, discard the particle exchange step. 

In order to carry out a fair comparison of the DPF and the centralized PF, the total number of particles must coincide. In the sequel, we present simulation results with several values of $M$, namely $M \in \left\lbrace 8, 16, 32, 64, 128 \right\rbrace$, while the number of particles per PE is kept fixed, $K=256$. For the centralized PF, the number of particles is set as $N=MK$, hence $N \in \{ 8\times K, 16\times K, 32\times K, 64\times K, 128\times K \}$. Note that the number of sensors collecting data, $J=18$, is kept fixed for all simulations, despite the variations in the number $M$ of PEs.

For each value of $M$ it is necessary to describe how the PEs are interconnected in order to carry out the particle exchanges specified by Algorithm \ref{alDPF}. These interconnections can be fully described by a simple graph, and hence we use the Havel-Hakimi algorithm \cite{Hakimi62} in order to generate them automatically for each $M$. The resulting graphs are such that every PE has exactly $M/4$ neighbors. The period of the exchange step is set to $n_0 = 10$, and at every exchange step each PE interchanges $\floor{3.6K/M}$ particles with every neighbor. Since each PE has $M/4$ neighbors, this amounts to approximately $90\%$ of the particles in each PE being swapped with particles belonging to its neighbors\footnote{A value of $90\%$ was chosen here to ensure that the aggregated weights of the PEs can be properly balanced even when one of them is much higher than the rest, i.e., when the particles of a single PE capture most of the importance weight. This proportion can be decreased, e.g., by reducing the period $n_0$ between exchange steps.}. The mapping $\beta$ that determines the particles to be exchanged is kept deterministic, but depends (in an obvious way) on the graph generated for each value of $M$.

We numerically assess whether Assumption \ref{asExchange} holds. The parameters involved are tentatively set to $c=4$, $q=4$ and $\epsilon=0.5$. According to Assumption \ref{asExchange}, for an arbitrary number of PEs, $M$, these values should yield an upper bound on the expectation of the supremum of the aggregated weights of the form
\begin{equation}
E\left[
	\left(
		\sup_{1 \le m \le M} W_{rn_0}^{(m)} 
	\right)^q
\right] \le \frac{c^q}{M^{q-\epsilon}}
\label{eqRepA3}
\end{equation}
after every exchange step (i.e., when $n=rn_0$, for $r \in \mathbb{N}$). 

\begin{figure}[htpb]
\begin{center}
	\includegraphics[width=0.8\linewidth]{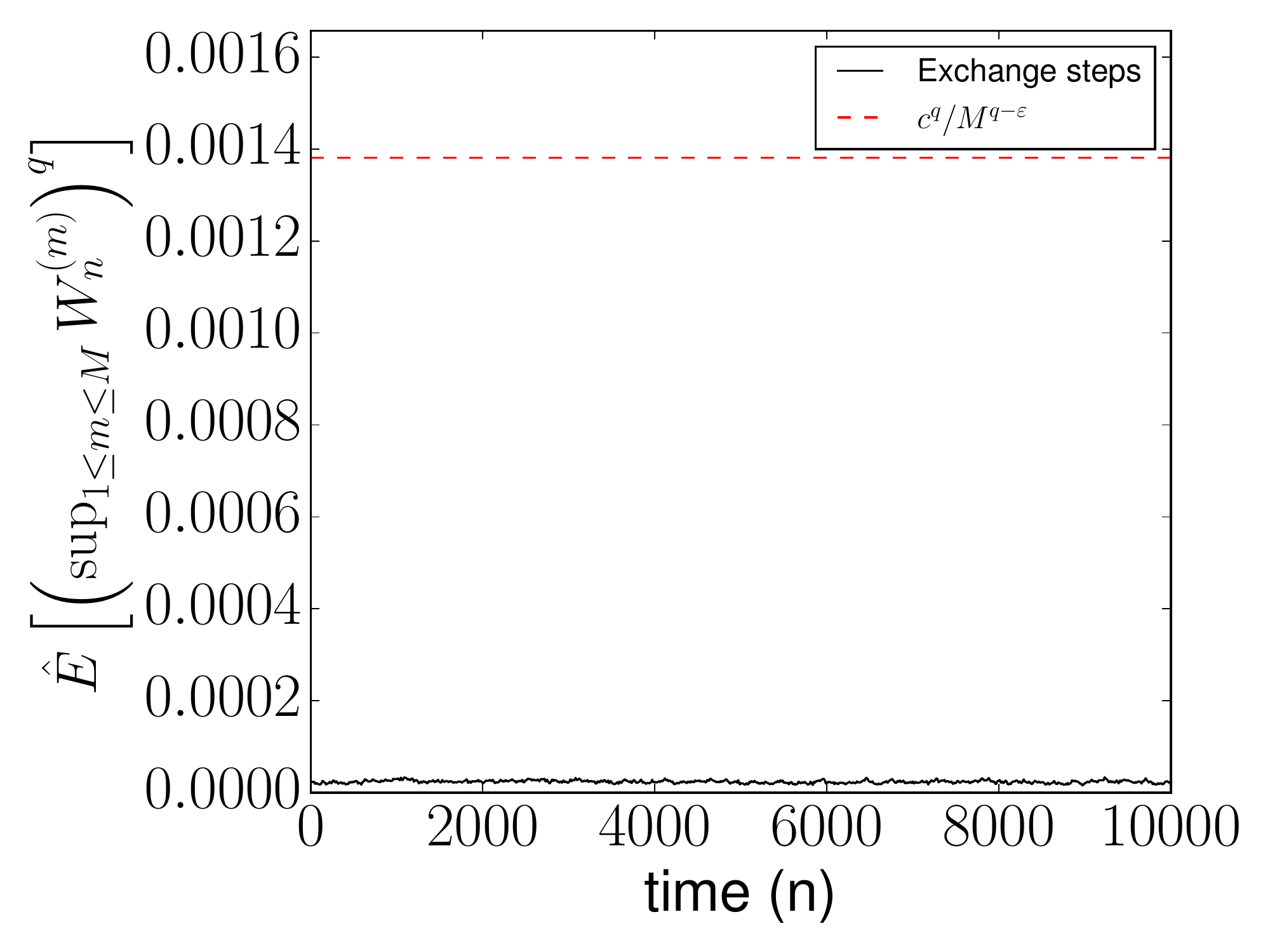}
\end{center}
\caption{Evolution over time of the supremum of the aggregate weights (to the fourth power) for $M=32$, averaged over 150 independent simulation runs, with constants $c=4$, $q=4$ and $\epsilon=0.5$. For clarity of visualization, only the time steps for which an exchange of particles is performed (i.e., $n=rn_0$) are shown.}
\label{f1b}
\end{figure}

In Figure \ref{f1b}, an estimate of the expectation on the left-had side (lhs) of \eqref{eqRepA3}, computed by averaging $150$ independent simulations, is plotted for every time instant, $n=rn_0\le 10,000$, along with the upper bound on the rhs of \eqref{eqRepA3} when $M=32$. It is clear from the figure that the estimate of the expectation in \eqref{eqRepA3} is well below the upper bound after every exchange step (but this is not necessarily the case at times steps $n \neq rn_0$, when exchange steps are not taken)

\begin{figure}[htpb]
\begin{center}
	\includegraphics[width=0.8\linewidth]{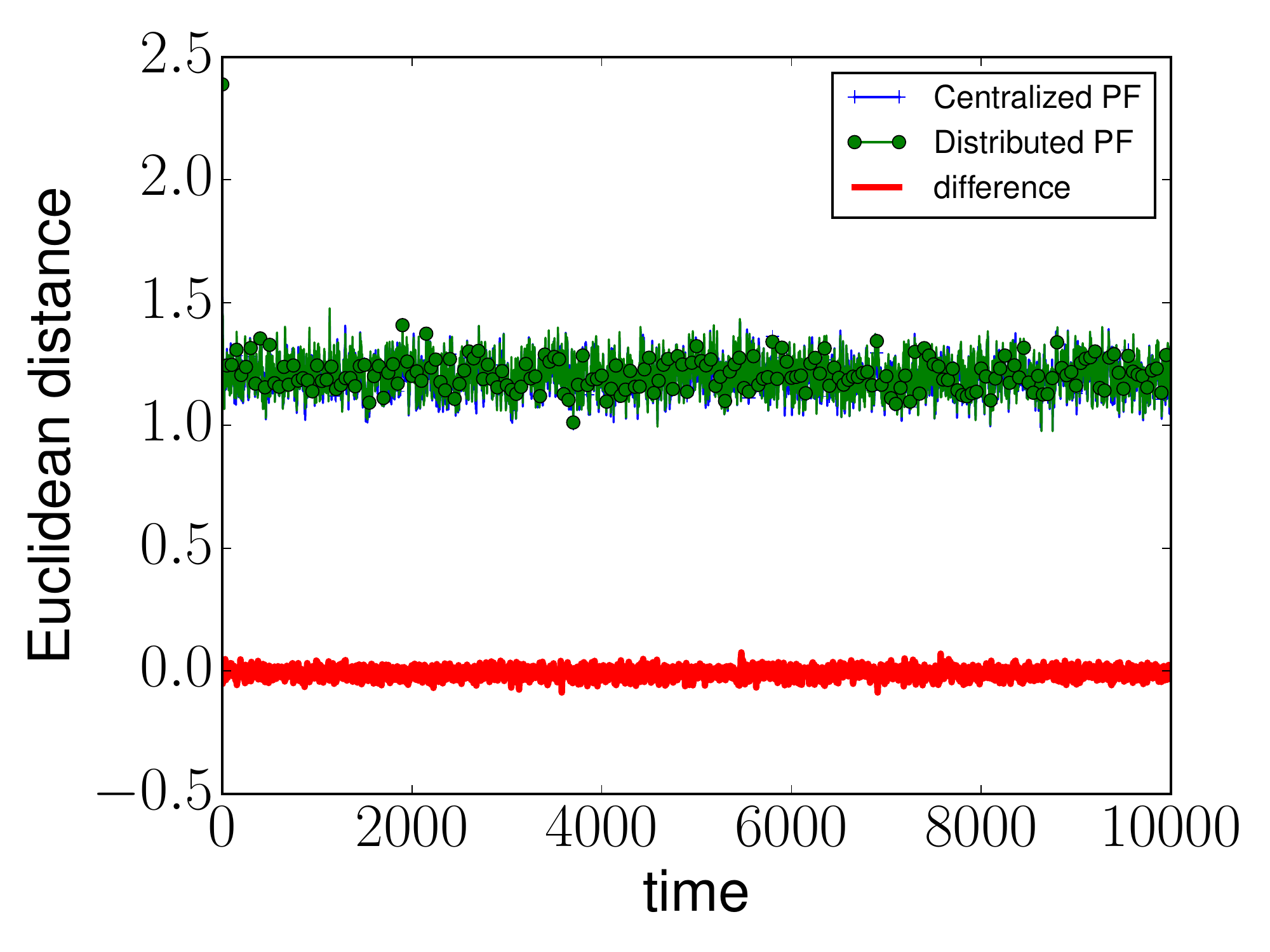}
\end{center}
\caption{Evolution over time of the $L_2$ errors (w.r.t. the true states) for the DPF with $M=32$, averaged over 150 independent simulation runs. The same errors for the centralized PF are also plotted. It is seen how the approximation error stays stable for up to $10,000$ times steps, as predicted by the uniform convergence result of Theorem \ref{thUniform1}.}
\label{f1a}
\end{figure}

Using the same set of 150 independent simulation runs, we have estimated the $L_2$ errors of the posterior mean of the state computed via the DPF algorithm (i.e., $\hat x_n^{MK} = \sum_{m=1}^M W_n^{(m)} \sum_{k=1}^K w_n^{(m,k)} x_n^{(m,k)}$) w.r.t. to the true value of the state signal $x_n$, for $1 \le n \le 10,000$. Figure \ref{f1a} shows the results. It is apparent that the error remains stable (it does not drift up) for the complete period of 10,000 time steps. Moreover, the performance is very close to the centralized PF with the same total number of particles, $N=MK=32\times 256$, for which the approximation errors are also show in Figure \ref{f1a} (note that the difference between the errors for the DPF and the errors for the centralized PF is also plotted).

\begin{figure}[htpb]
\begin{center}
	\includegraphics[width=0.8\linewidth]{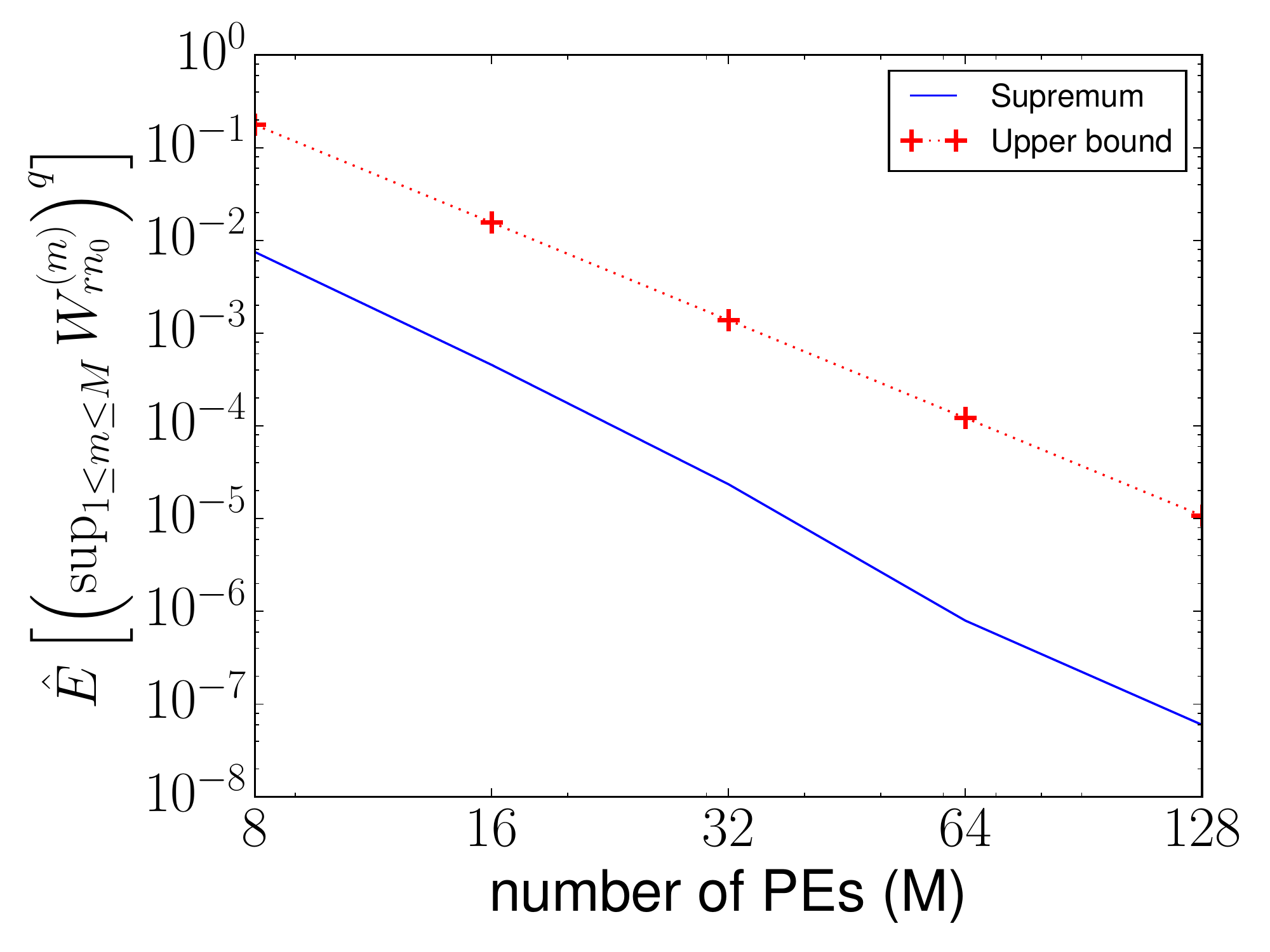}
\end{center}
\caption{Evolution of the expected supremum of the aggregate weights (to the fourth power, with constants $c=4$, $q=4$ and $\epsilon=0.5$) as a function of the number of PEs, $M$, for a fixed time instant $n=rn_0=1,000$. The expectation is estimated from a set of 230 independent simulation runs for each value of $M$. The upper bound prescribed by Assumption \ref{asExchange} is plotted as a dashed line.}
\label{f2a}
\end{figure}

We have carried out additional computer simulation trials with $M=8, 16, 32, 64$ and $128$ in order to verify whether the parameter set $\{ c=4, q=4, \epsilon=0.5\}$ appears to be independent of $M$, as demanded by Assumption \ref{asExchange}. For each value of $M$, we have run 230 independent computer simulations with $n = 1, \ldots, 3,000$ time steps. Figure \ref{f2a} depicts, for fixed $n=100n_0=1,000$ and $M=8, 16, 32, 64, 128$ the (estimated) expectation and upper bound that correspond, respectively, to the lhs and the rhs of \eqref{eqRepA3}. It can be observed that the expectation decreases, along with the upper bound, as $M$ increases. However, the ratio between the bound on the rhs of \eqref{eqRepA3} and the expectation on the lhs of \eqref{eqRepA3} becomes larger as $M$ is increased: for this set of simulations, it ranges from $\approx 22$ when $M=8$ to $\approx 1.25 \times 10^3$ for $M=128$. These numerical results strongly suggest that Assumption \ref{asExchange} holds true for this particular example.

\begin{figure}[htpb]
\begin{center}
	\includegraphics[width=0.8\linewidth]{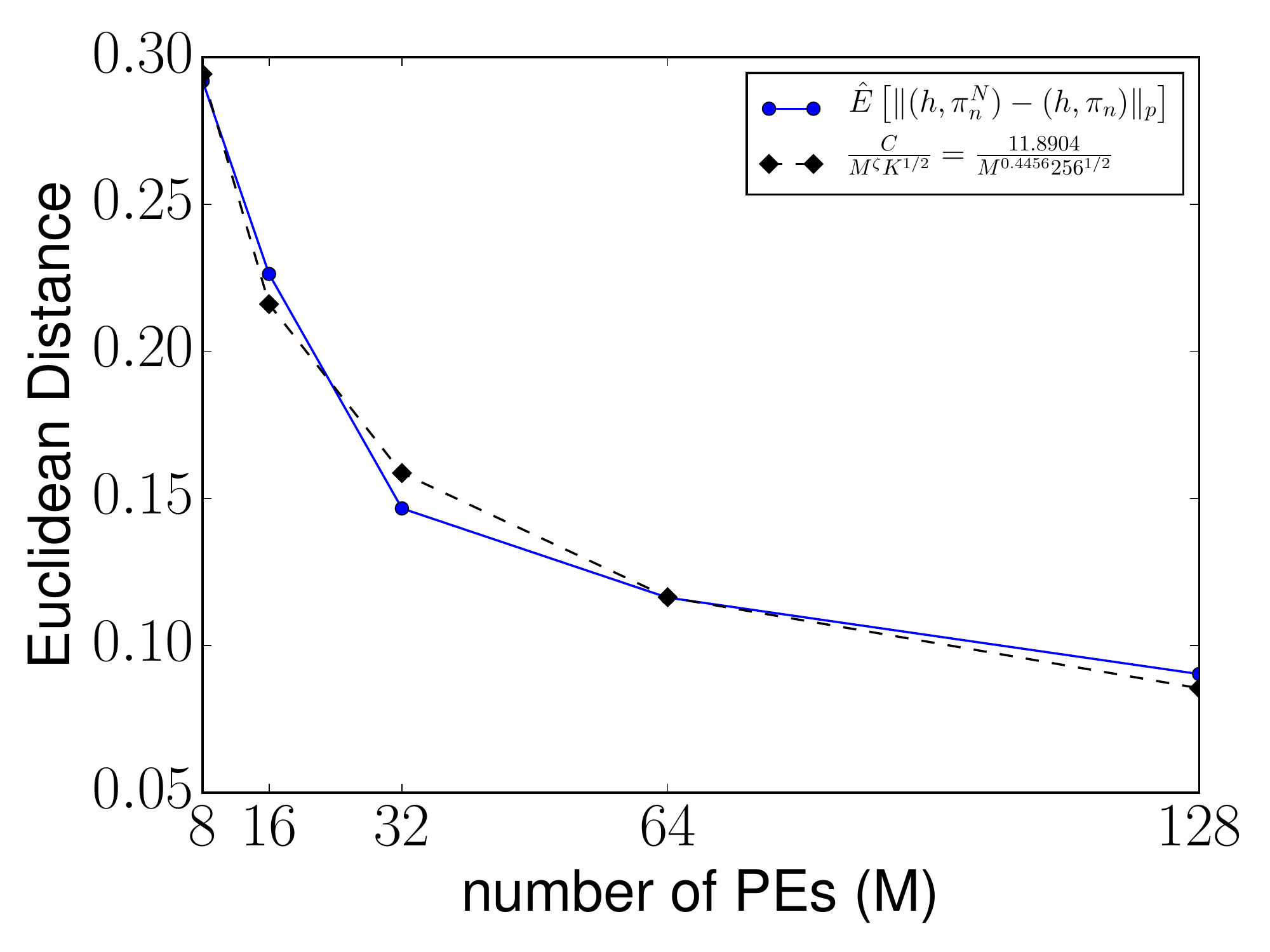}
\end{center}
\caption{Approximate $L_2$ errors of the position estimates for the DPF for different values of $M$. An exponentially decreasing function whose parameters are fitted by least squares using the empirical $L_2$ errors is also displayed.}
\label{f2b}
\end{figure}

Finally, we aim at evaluating the rate at which the $L_2$ errors in the approximation of the posterior mean of the state converge with increasing $M$. To compute these errors, since the true posterior mean of the state, namely the integral $\hat x_n = \int x_n \pi_n(dx_n)$, cannot be computed exactly, we have used the estimates provided by a centralized PF with $N=MK=128\times 256 = 2^{15}$ particles as a proxy for the actual $\hat x_n$. Then, using the same set of 230 independent simulations as in Figure \ref{f2a}, we have empirically estimated the $L_2$ errors for $M=8, 16, 32, 64, 128$ and $n=2,000$, and plotted them in Figure \ref{f2b}. To obtain an empirical convergence rate, we have used the obtained $L_2$ errors to fit an exponentially decreasing function of the form $\frac{C}{M^\zeta N^\frac{1}{2}}$, where $C$ and $\zeta$ are constants. The result, using a least squares fit, is $C \approx 11.8$ and $\zeta \approx 0.44$, which is close to the optimal Monte Carlo rate of $M^{-\frac{1}{2}}$.


\section{Conclusions} \label{sConclusions}

We have introduced the first rigorous proof of convergence for a particle filter (PF) based on the popular distributed resampling with non-proportional allocation (DRNA) scheme of \cite{Bolic05}. We have provided sufficient conditions for the uniform convergence of the resulting distributed PF over time. Explicit error rates in terms of the number of processing elements (PEs) and the number of particles per PE have been obtained. Uniform convergence guarantees that the distributed PF can be run for an arbitrarily long sequence of observations without requiring to increase the computational load over time. This kind of convergence is inherently stronger than the consistency proofs in classical papers such as \cite{Crisan02} as well as in more recent contributions like \cite{Lee13,Miguez13b,Crisan14a}. As for future work, we believe that some recently developed theoretical techniques \cite{Whiteley13} could be applied in order to relax some of the assumptions made for the analysis and/or to improve on the error rates found in this paper.

In order to corroborate the validity of the analysis and to assess the practical performance of the distributed algorithm, we have carried out computer simulations for an indoor target tracking problem. 
The assumptions on which our analysis relies are standard in the literature for centralized PFs \cite{DelMoral01c}, except for Assumption \ref{asExchange} that is needed to handle the particle exchange scheme and is key to prove convergence, therefore we have deveoted most of the computer simulation study to show that it holds numerically. We have also compared the position estimation error attained by distributed PF of interest and a standard (centralized) PF, and found that the two algorithms display a very similar performance. 

\section*{Acknowledgments}

This work was supported by {\em Ministerio de Econom\'{\i}a y Competitividad} of Spain (project COMPREHENSION TEC2012-38883-C02-01), {\em Comunidad de Madrid} (project CASI-CAM-CM S2013/ICE-2845) and the Office of Naval Research Global (award no. N62909- 15-1-2011). A preliminary version of this paper was presented at the IEEE Sensor, Array and Multichannel Signal Processing Workshop (SAM) 2014.

At the time of the original submission, J. M. was with Departamento de Teor\'{\i}a de la Se\~nal y Comunicaciones, Universidad Carlos III de Madrid.

\appendix

\section{Proof of Lemma \ref{lmW}} \label{apLemmaW}

Let us denote $W_n^M = \sup_{1 \le m \le M} W_n^{(m)}$ for conciseness. We follow the same type of argument as in the proof of \cite[Lemma 4.1]{Crisan14a}. Choose a constant $\gamma$ such that $\epsilon < \gamma < q-1$ and define
\begin{equation}
U_n^{\gamma,q} = \sum_{M=1}^\infty M^{q-1-\gamma} (W_n^M)^q.
\label{eqShow_it_is_measurable}
\end{equation}
The random variable $U_n^{\gamma,q}$ is obviously non-negative and, additionally, it can be shown that it has a finite mean, $E[U_n^{\gamma,q}]<\infty$. Indeed, from Fatou's lemma
\begin{equation}
E\left[
        U_n^{\gamma,q}
\right] \le \sum_{M=1}^\infty M^{q-1-\gamma} E[(W_n^M)^q]
\le c^q \sum_{M=1}^\infty M^{-1-\gamma+\epsilon}, \label{eqSerie}
\end{equation}
where the second inequality follows from Eq. \eqref{eqAss_lema_W} in the statement of Lemma \ref{lmW}. Since $\gamma-\epsilon>0$, it follows that $\sum_{M=1}^\infty M^{-1-(\gamma-\epsilon)} < \infty$, hence $E[U_n^{\gamma,q}]<\infty$.

We use the so-defined r.v. $U_n^{\gamma,q}$ in order to determine the convergence rate of $W_n^M$. Obviously, $M^{q-1-\gamma} \left( W_n^M \right)^q \le U_n^{\gamma,q}$ and solving for $W_n^M$ yields
$
W_n^M \le \frac{
        \left(
                U_n^{\gamma,q}
        \right)^\frac{1}{q}
}{
        M^{1 - \frac{1+\gamma}{q}}
}.
$
If we define $\varepsilon = \frac{1+\gamma}{q}$ and $U_n^\varepsilon = \left( U_n^{\gamma,q} \right)^\frac{1}{q}$, then we obtain the inequality
$
W_n^M \le \frac{U_n^\varepsilon}{M^{1-\varepsilon}}.
$
Since $E[U_n^{\gamma,q}] < \infty$, it follows that $E\left[ \left( U_n^\varepsilon )^q \right) \right] < \infty$, hence $U_n^\varepsilon$ is a.s. finite. Also, we recall that $0 \le \epsilon < 1$ and $\epsilon < \gamma < q-1$, therefore $\frac{1+\epsilon}{q} < \varepsilon < 1$. 


Finally, note that 
$$
E\left[ \left( U_n^\varepsilon )^q \right) \right] = E[U_n^{\gamma,q}] < c^q \sum_{M=1}^\infty M^{-1-\gamma+\epsilon},
$$
independently of $n$, as shown by \eqref{eqSerie}, hence it is enough to choose $u^{\varepsilon,q} = c^q \sum_{M=1}^\infty M^{-1-\gamma+\epsilon} < \infty$ in order to complete the proof. $\qed$

%
\section*{References}

\footnotesize
\bibliographystyle{plain}
\bibliography{bibliografia}

\newpage
\pagestyle{empty}

\newpage

\newpage

\newpage

\newpage

\end{document}